\documentclass[preprint,journal]{vgtc}           

\onlineid{0}

\vgtccategory{Research}

\vgtcpapertype{application/design study}

\title{SpeechMirror: A Multimodal Visual Analytics System for Personalized Reflection of Online Public Speaking Effectiveness
}

\author{
  \authororcid{Zeyuan Huang}{0000-0002-2639-0487}, 
  \authororcid{Qiang He}{0009-0006-2140-1681}, 
  \authororcid{Kevin Maher}{0000-0002-6486-4866}, 
  \authororcid{Xiaoming Deng}{0000-0001-8576-1201}, \textit{Member, IEEE}, 
  \authororcid{Yu-Kun Lai}{0000-0002-2094-5680}, \textit{Member, IEEE}, \\
  \authororcid{Cuixia Ma}{0000-0003-3999-7429}, 
  \authororcid{Sheng-feng Qin}{0000-0001-8538-8136}, 
  \authororcid{Yong-Jin Liu}{0000-0001-5774-1916}, \textit{Senior Member, IEEE}, 
  and \authororcid{Hongan Wang}{0000-0002-9320-4920}, \textit{Member, IEEE}}

\authorfooter{
    \item
     Z.Y. Huang, Q. He, X.M. Deng, C.X. Ma, and H.A. Wang are with the 
     Beijing Key Laboratory of Human-Computer Interaction, Institute of Software, Chinese Academy of Sciences and also with the School of Computer Science and Technology, University of Chinese Academy of Sciences. E-mail: \{zeyuan2020, heqiang2022, xiaoming, cuixia, hongan\}@iscas.ac.cn.

    \item
    Kevin Maher is with Diatom Design Limited Liability Company. E-mail: kevintmaher@gmail.com.

    \item
    Y.-K. Lai is with
    the School of Computer Science and Informatics,
    Cardiff University. E-mail: LaiY4@cardiff.ac.uk.
    
    \item
    S. Qing is with the School of Design, Northumbria University. E-mail: sheng-feng.qin@northumbria.ac.uk.
    
    \item
    Y.-J. Liu is with the Department of Computer Science and Technology, MOE-Key Laboratory of Pervasive Computing, Tsinghua University. E-mail: liuyongjin@tsinghua.edu.cn.
    \item 
    C.X. Ma, Y.-J. Liu, H.A. Wang are corresponding authors. 
}

\abstract{
As communications are increasingly taking place virtually, the ability to present well online is becoming an indispensable skill. Online speakers are facing unique challenges in engaging with remote audiences. However, there has been a lack of evidence-based analytical systems for people to comprehensively evaluate online speeches and further discover possibilities for improvement. This paper introduces SpeechMirror, a visual analytics system facilitating reflection on a speech based on insights from a collection of online speeches. The system estimates the impact of different speech techniques on effectiveness and applies them to a speech to give users awareness of the performance of speech techniques. A similarity recommendation approach based on speech factors or script content supports guided exploration to expand knowledge of presentation evidence and accelerate the discovery of speech delivery possibilities. SpeechMirror provides intuitive visualizations and interactions for users to understand speech factors. Among them, SpeechTwin, a novel multimodal visual summary of speech, supports rapid understanding of critical speech factors and comparison of different speech samples, and SpeechPlayer augments the speech video by integrating visualization of the speaker's body language with interaction, for focused analysis. The system utilizes visualizations suited to the distinct nature of different speech factors for user comprehension. The proposed system and visualization techniques were evaluated with domain experts and amateurs, demonstrating usability for users with low visualization literacy and its efficacy in assisting users to develop insights for potential improvement.
} 


\graphicspath{{figs/}{figures/}{pictures/}{images/}{./}} 

\usepackage{tabu}              
\usepackage{booktabs}           
\usepackage{lipsum}            
\usepackage{mwe}                       

\usepackage{mathptmx}                
\usepackage{multirow}
\usepackage{enumitem}

\begin{document}

\firstsection{Introduction}

\maketitle

In recent years there is a rapidly growing trend of virtual communication. 
For speakers, the transition from offline to online opens up new ways to express ideas. However, it also poses challenges for speakers to engage audiences at a distance. 

Guidance for better virtual presentations has been provided for general tips \cite{presentingvirtually2021}, workplace meetings \cite{potter2022presenting}, and various virtual scenarios in remote education \cite{green2021presenting}. 
However, these tips only give theoretical advice that 
can be difficult to be used to evaluate or guide practical presentation delivery.
For speaking there are inconsistencies between different theories about speech techniques \cite{multimodalted2020, effective2022}.
It is time consuming and experience demanding for speakers to understand how their speech performs and even harder 
to find references for potential improvement.

Existing literature has reported research on the analysis of speech presentation techniques.
Some studies have analyzed the relationship between individual \cite{upperbodygesture2012, languagecharacteristics2014, nonverbalbehavioralcues2015, areyourtedtalk2015, awetheaudience2018} or multiple \cite{estimationvideokinect2014,estimationslideaudio2014,multimodalassessment2015,deepmultimodalfusion2016,evaluatingspeech2015,multichannel2018} speech factors and the effectiveness of the speech. 
However, while these methods typically 
analyze
effectiveness of speech factors in the form of scores, it is challenging for users to grasp the underlying reasons for the score and consequently identify areas for improvement.
Visual analytics systems have been proposed to facilitate an interactive exploration of presentation techniques. 
Existing visual analytics methods can be classified into two dimensions: single factor versus multiple factors, and individual speech versus a collection of speeches.
However, these works are unable to directly assess the effectiveness of speech factors, nor do they provide a comprehensive list of speech factors for analysis.
Although existing works mainly rely on video inputs, none of them are specifically designed for online public speaking scenarios. Connecting with remote audiences through a camera requires some particular presentation techniques that deserve further exploration.

In this work, we propose \textit{SpeechMirror}, a visual analytics system that allows experts and amateurs in public speaking to gain insights into a speech driven by a large-scale analysis of online speeches. \textit{SpeechMirror} can help understand areas for improvement and recommend examples of speeches as references for practicing improvement.

The system allows enhanced ability to understand the estimated effectiveness of different techniques in a speech. 
While various ideas exist for measuring speech effectiveness \cite{effective2022}, we utilize a collection of videos ranked in a speech contest as our quantifiable metric.
To understand the underlying techniques that influence the effectiveness of online speeches, we establish various multimodal speech factors from domain interviews and existing literature. 
We especially consider techniques that are different in online speeches including the use of stage, eye contact, and body gestures.
Based on our collection of speeches, we determine the relationship between various techniques and effectiveness. With these identified relationships, they can be applied to detect the areas for improvements (via diagnostics) and recommend some techniques via examples to improve the speech effectiveness (via prediction). Thus, users can gain awareness of the estimated impact of various factors on a speech as a whole or in individual sentences. 

\textit{SpeechMirror} offers a personalized recommendation approach to explore various possibilities for speech delivery. The recommendation is based on a selected part of speech and produces results at different granularity levels (in entire speech or individual sentence) and different modes (by factors or script contents). The recommendation allows reflection on a speech compared with other speeches in the collection, which expands user knowledge of possible expressions.

\begin{figure*}[ht!]
 \setlength{\belowcaptionskip}{-0.6cm}
  \setlength{\abovecaptionskip}{0.2cm}
 \centering 
  \includegraphics[width=\linewidth]{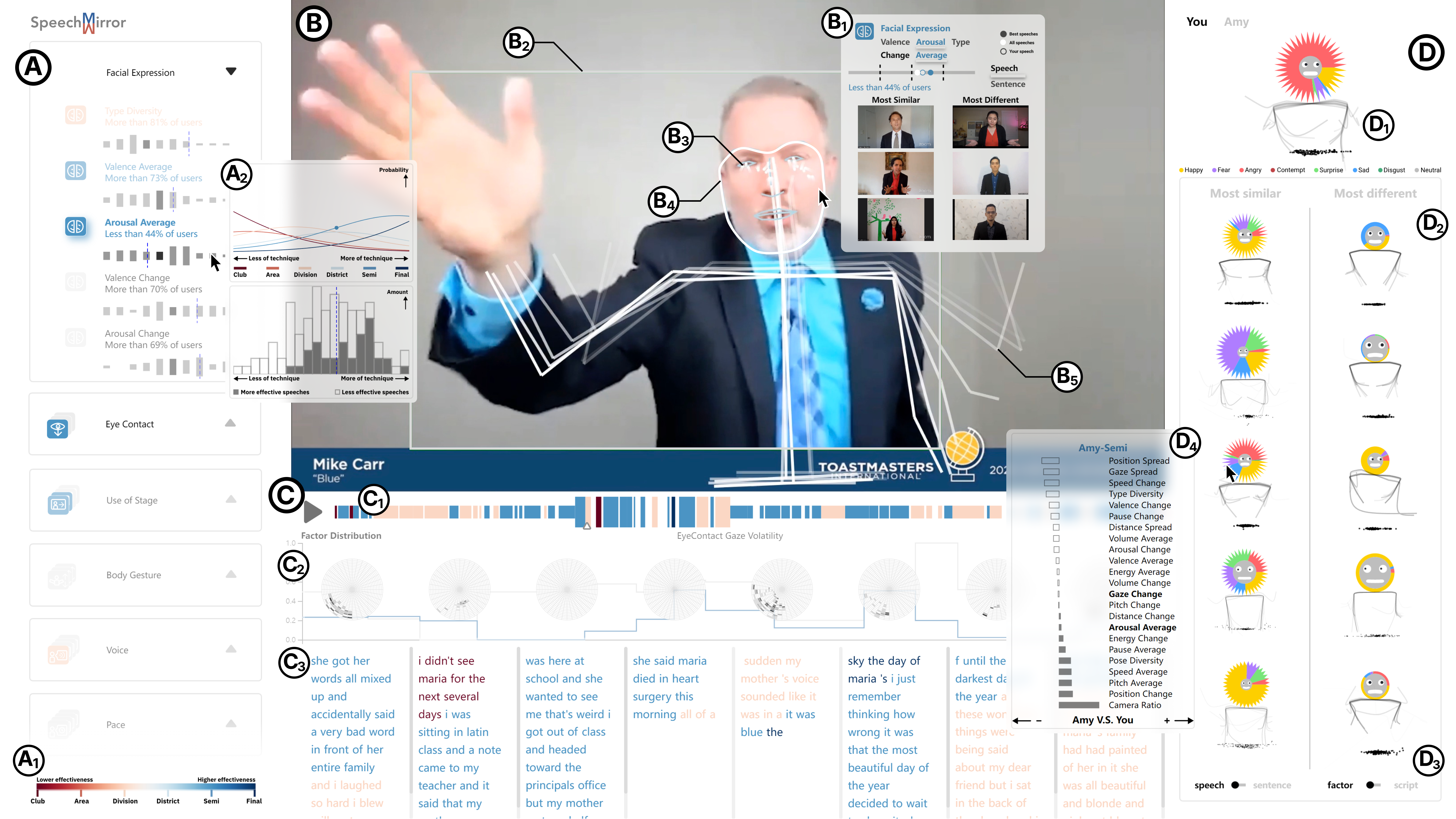}
 \caption{Our visual analytics system supports the evaluation and understanding of presentation techniques as well as discovering expression possibilities. The Factor Panel (A) delivers feedback on the effectiveness of speech factors and assists in comprehending the trends of speech factor effectiveness. The Speaker Panel (B) provides an augmented video integrating visualizations of presentation techniques. The Time Slice Panel (C) enables users to understand the effectiveness and factor data with script over time. The Mirror Panel (D) recommends similar or different speeches and facilitates intuitive comparison between speeches with visual summary. }
  \label{fig:teaser}
\end{figure*}

Challenges to the scope of the interface of our system include the complexity brought in by a number of multimodal factors, a large amount of changes of the factors over time, as well as supporting comparison and contrast of different speech samples. Given the low visualization literacy of our target users, we provide intuitive visualizations for better understanding. Specifically, we design SpeechTwin, a novel visual speech summary, to facilitate rapid comprehension of speech factors and comparison between speeches. In order to allow an enhanced understanding of body language in the original context, SpeechPlayer directly visualizes presentation techniques in the video feed. Also significant is a series of different glyphs developed to support an intuitive understanding of speech features.

The major contributions of our work include: 
(1) a visual analytics system to assist in analyzing the effectiveness of an individual online speech within a collection of speeches;
(2) novel visualization designs integrating multimodal features and factors to support the understanding of the use of presentation techniques;
(3) usability of our system demonstrated and analytical insights gained in an evaluation study.

\section{Related Work}

\subsection{Automated Analysis of Public Speaking}

With the assistance of computers, presentations are digitalized and analyzed in a quantitative way. The research can be classified into two major directions in this discussion.

\textbf{Computation-centered approaches} focus on building computational models to evaluate the effectiveness and significance of different presentation techniques. Some research works focused on analyzing presentation techniques in single modality, such as upper body gesture \cite{upperbodygesture2012}, language characteristics  \cite{languagecharacteristics2014}, body movements \cite{nonverbalbehavioralcues2015}, prosodic voice characteristics \cite{areyourtedtalk2015}, narrative trajectories \cite{awetheaudience2018}, etc. 
Additional efforts have been made to take advantage of multimodal information for analysis. To estimate the presentation performance level, Echeverría et al. \cite{estimationvideokinect2014} used eye contact from video, body posture, and body language from a Kinect, and Luzardo et al. \cite{estimationslideaudio2014} used slides and audio. Wörtwein et al. \cite{multimodalassessment2015} assessed public speaking performance with audiovisual features. Nojavanasghari et al. \cite{deepmultimodalfusion2016} studied persuasiveness prediction by a deep multimodal fusion method with visual, acoustic, and text descriptors. Ramanarayanan et al. \cite{evaluatingspeech2015} scored presentations with speech, face, emotion and body movement.
Sharma et al. \cite{multichannel2018} predicted video popularity of TED (technology, entertainment and design) talks from facial and physical appearance, facial expressions, and pose variations.
These efforts 
evaluate speech performance based on indicators of effectiveness such as video popularity and audience ratings. In our work, we adopted a speech contest scenario, with the contest placement serving as a measure of speech effectiveness. 
With a computation-based approach, it is easy for speakers to obtain performance assessment, but hard to further investigate why and how to do better. 
In our system, we provide insights on speech factor effectiveness and further support exploration for areas of improvement. 

\textbf{Visualization-centered approaches} present the data of presentation techniques through visualizations and further support users in gaining insights through exploration. 
Existing literature focuses on different purposes of speech analysis: identifying narration strategies \cite{speechlens2019}, understanding emotion \cite{emoco2020}, training vocal skills \cite{voicecoach2020}, analyzing debate transcripts \cite{debatevis2020}, exploring presentation techniques \cite{multimodalted2020}, decomposing humor \cite{dehumor2021}, assisting the exploration of gestures \cite{zeng2022gesturelens}, and analyzing the effectiveness of different speech factors \cite{effective2022}. While prior works have provided approaches to understand the contributions of speech factors or recommend evidence for public speaking training, our work aims at evaluating a single speech, assisting users to analyze the speech in the 
context of a
speech collection.

Current research mainly focuses on offline presentations or taking a video as input. 
Distinct from in-person presentations, there are unique techniques that can make online presentations more effective.
To the best of our knowledge, there is still a lack of qualitative research in the online public speaking scenario.

\subsection{Online Public Speaking}

The importance of online public speaking has increased in recent years. The consulting firm Gartner predicts that by 2024 only 25\% of business meetings are offline \cite{rimol2021gartner}. 
In the testing phase of a system including both online and offline speeches, an experienced public speaking expert reasoned that there were significant differences in the emotional expression of online and offline contests \cite{effective2022}.
Teodorescu et al. \cite{teodorescu2020experimental} found possible differences in NLP (Natural Language Processing) of titles in an online and offline contest, focused on microcontroller applications. 

There are other important differences of online and offline speeches that could be uniquely investigated by analytical systems. Since the fixed reference frame of the camera and fixed position of the microphone are perceived the same to all members of the audience, the effects of eye contact, stage movement, and voice volume can be better measured. Ochoa et al. created an automated system measuring speech delivery to a remote audience \cite{ochoaRAP}.

\subsection{Human Physical Behavior Data Visualization}
\label{sec:human-physical-behavior}
Our work focuses on body language, including facial expression, eye gaze, gestures, and positional movement. 

Gesture and positional information is important to understand in speeches. 
In sports, Stein et al. \cite{stein2018Sports} integrated movement data including possible player movement directly with the original video, ``enabling analysts to draw on the advantages of both''. Mova \cite{mova2014alemi} used small multiples
of extracted body keypoints that were annotated with color to indicate features of movement. A design prototype we considered included these enhanced keypoints. In public speaking, GestureLens \cite{zeng2022gesturelens} provided interactions to support the understanding of gestures in relation to content and time. In contrast to their work, we focus on how to show gestures not 
only between different words in a sentence, but also creating representative skeletons for custom time ranges.

There are many strategies to visualize eye gaze. Much research focused on how to visualize where gaze is directed, such as fixation points and saccades \cite{d2018eye}, and heatmaps \cite{pakov2007visualization}. Our work instead focuses on the direction that a speaker is looking at, which can more clearly show eye contact or gaze directions that fit with speech content. Similarly, Higuch et al. developed visualizations to show the eye gaze direction of autistic children \cite{higuchi2018visualizing}.  

The complexity of facial expression has led to a variety of visualization strategies to present them. E-ffective used a spiral visualization to show emotional shifts in speakers as well as a text-based visualization to display emotion in the context of a speech script \cite{effective2022}. In EmotionCues, Zeng et al. created an Emotion Band, a visualization that allows users to track emotion changes of multiple users in a flow-based design \cite{emotioncues2020}. For our work we were interested in presenting the relative differences for a range of emotional factors, and developed visualizations for a range of these factors.

\section{Design of SpeechMirror}

In this section, we will introduce the domain-centered design of \textit{SpeechMirror}. The design process mainly contains two stages: literature reviews and domain interviews with public speaking experts and amateurs. With the insights gained, we derived the design considerations and the scope of presentation techniques in our system.

\subsection{Design Process}
The main goal of our work is to assist public speaking experts and amateurs in understanding a speech by means of guided insights from a collection of speeches. 
We first reviewed the existing literature to build a list of speech factors that are potentially related to the effectiveness of public speaking. We also developed an understanding of the distinct characteristics of online speech presentations.

To better understand the demands amateurs and experts valued as important,  
we conducted 30-minute semi-structured interviews with 6 volunteer participants, including 4 experts (DE1 - DE4) and 2 amateurs (DA1 - DA2). Both the amateurs and the experts had participated in the World Championship of Public Speaking (WCPS) contest at least once. The experts all had experience coaching public speaking as an occupation. 
The amateurs have experience in online public speaking contests between 2-5 times and have watched no more than 50 speeches.

The goal of the interviews is to understand their opinions on effective online presentation techniques, current practice and challenges in developing public speaking skills, as well as their needs for a new speech video analysis tool. Detailed information on the interviews is provided in Sect. 1 of the supplemental material.

\subsection{Design Considerations}
\label{sect:design-considerations}

We derived the design considerations for \textit{SpeechMirror} from existing literature and our domain interviews.

\textbf{DC1: Provide an integrated way of understanding the effectiveness of speech factors in a speech within a speech collection.}
It is important to reveal the effectiveness of speeches. 
In our interviews, DE4 mentioned the requirement of having a benchmark to measure speech performance. 
Focusing on providing exploration of factors within a speech collection, E-ffective showed both the importance and difficulty in understanding the effectiveness of speech factors through the feedback ratings. 
To migrate the challenges, we sought to provide users' insights on their speeches in \textit{SpeechMirror}, so that users can understand the speech performance.
In the interviews, DE3, a full-time public speaking trainer, thought there are many factors in speeches that trainers need to keep in mind for feedback. He remarked, ``\textit{a trainer might ignore or forget the top 8 (factors), but 
a system might be able to tell the top 9.}''
Thus we sought out a more comprehensive set of effectiveness factors for consideration, as well as a more integrated way for understanding the factors.

\textbf{DC2: Provide a convenient approach for obtaining speeches for reference.}
A promising study \cite{voicecoach2020} demonstrated the potential for improvement of speeches by discovery of high quality examples. 
They claimed the future improvement  of showing negative samples as warnings to avoid.
The importance of references was also brought to our attention during our interview sessions. 
In our interviews with less experienced amateurs, they voiced needs to discover different ways of expression. When discussing the use of body language, DA2 stated ``\textit{a problem is that I only compare to what I know. With the same (speech) content, I wonder how others express it.}'' DA1 similarly expressed uncertainty about the use of eye contact and gestures.
More generally, DE2 claimed priority should be placed in video referencing, ``\textit{study the people who do better than you are.}'' 
Our system aims to offer speech samples with similar or different speech delivery or content to facilitate exploration. These samples exhibit varying levels of effectiveness to allow the comprehension of possible expressions.

\textbf{DC3: Reveal the temporal distribution of the effectiveness of speech factors.}
The order of techniques in speeches matters, and many sources have theories about advantages of using techniques at different times. The WCPS 2012 champion advised stage movement where the speaker should ``end your speech in the same location where you began'' \cite{speakerleaderchampion}. A university textbook on public speaking claims the order of gestures as they appear in relation to the main idea in the speech can be important \cite{mclean2015business}.
Our interviews also revealed opportunity for speakers to learn from time order. DE3 claimed speakers might not only be unclear about what they should do in their speech, but also ``\textit{not be clear what they \textit{did} in their speech. Did they smile at the beginning?}'' \textit{SpeechMirror} supports the needs of understanding the use of presentation techniques over time in an intuitive way.

\textbf{DC4: Demonstrate speech factors in relation to the multimodal context within a speech.}
Our interviews revealed the necessity of viewing speech factors, such as the factor raw data and factor effectiveness, in reference to the broader context of the speech. In our interview DE4 stated that ``\textit{content is the most important, especially cultural background}'', pointing out the significance of verbal context. Displaying how non-verbal techniques relate to the verbal context is commonly used in visual analytic systems for speech analysis, e.g. \cite{emoco2020, multimodalted2020, dehumor2021, zeng2022gesturelens, effective2022}. The video context was also claimed to be important in our interviews. DE3, DE2, DA1 stated that for online speeches, understanding how to move in the limited space was crucial.  Other interviewees brought up factors such as eye contact, stage movement, and using gestures within the limited space, all of which are difficult to understand without reference to the original video. Thus, we intend to provide the original video context and verbal context to enhance understanding of factors inside a speech.

\textbf{DC5: Summarize the most relevant techniques used.}
We evaluate 23 different techniques in speeches. Given the large amount of information at hand, visualization methods that provide relevant summaries of the techniques are crucial. The data of the techniques are time series, so it is hard for users to grab key information within the data. Data aggregation and abstraction are thus necessary.
Relevancy for the summaries both depends on the significance of the factor and the different use of the technique in terms of speeches in the collection. Less significant information is deemphasized or hidden from view. 
The summaries aim to enhance browsing of speeches, provide key information about the user's speech, and assist comparison of different speeches.

\begin{table}[t]
\caption{Presentation techniques, multimodal features and corresponding factors considered in \textit{SpeechMirror}. 
The significance of factors with * indicates a significant correlation with speech effectiveness ($p < 0.05$).
}
\vspace{-0.2cm}
\label{tab:pretech}
\resizebox{\columnwidth}{!}{
\begin{tabular}{llll}
\hline
Presentation Technique             & Feature                               & Factor                & Significance                        \\ \hline
\multirow{5}{*}{Facial Expression} & Type                                  & Diversity             & $p=0.002^*$      \\ \cline{2-4} 
                                   & \multirow{2}{*}{Valence}              & Volatility            & $p=0.571$          \\
                                   &                                       & Average               & $p=0.005^*$      \\ \cline{2-4} 
                                   & \multirow{2}{*}{Arousal}              & Volatility            & $p=0.761$              \\
                                   &                                       & Average               & $p=0.000^*$      \\ \hline
\multirow{3}{*}{Eye Contact}       & \multirow{2}{*}{Gaze Direction}             & Volatility            & $p=0.002^*$       \\
                                   &                                       & Dispersion            & $p=0.067$ \\ \cline{2-4}
                                   & Watching Camera     & Ratio    & $p=0.265$   \\ \hline
\multirow{4}{*}{Use of Stage}      & \multirow{2}{*}{Distance from Camera} & Volatility            & $p=0.908$             \\
                                   &                                       & Dispersion            & $p=0.185$             \\ \cline{2-4} 
                                   & \multirow{2}{*}{Position in Frame}    & Volatility            & $p=0.026^*$      \\
                                   &                                       & Dispersion            & $p=0.141$      \\ \hline
\multirow{3}{*}{Body Gesture}      & \multirow{2}{*}{Gesture Energy}       & Volatility            & $p=0.860$             \\
                                   &                                       & Average               & $p=0.426$             \\ \cline{2-4} 
                                   & Gesture Diversity                     & Diversity             & $p=0.266$      \\ \hline
\multirow{4}{*}{Voice}             & \multirow{2}{*}{Volume}               & Volatility            & $p=0.000^*$             \\
                                   &                                       & Average               & $p=0.413$ \\ \cline{2-4} 
                                   & \multirow{2}{*}{Pitch}                & Volatility            & $p=0.438$             \\
                                   &                                       & Average               & $p=0.988$             \\ \hline
\multirow{4}{*}{Pace}              & \multirow{2}{*}{Speaking Rate}        & Volatility            & $p=0.617$             \\
                                   &                                       & Average               & $p=0.198$             \\ \cline{2-4} 
                                   & \multirow{2}{*}{Pauses}               & Volatility            & $p=0.157$             \\
                                   &                                       & Average               & $p=0.533$             \\ \hline
\multirow{1}{*}{Content}           & Script                                &          \multicolumn{1}{c}{-}             &               \multicolumn{1}{c}{-}                      \\ \hline
\end{tabular}
}
\vspace{-0.6cm}
\end{table}

\subsection{Scope of Presentation Techniques}

There are many techniques that go into successful public speaking. 
Schneider et al. \cite{PresentationTrainer2017} provide an extensive list collected from public speaking experts about non-verbal communication practices. 

For online speaking there are additional techniques to consider, due to the limited camera view, single viewing angle of the audience, etc. Several techniques  critical to effective online speaking have yet to be intentionally visualized for further analysis.
Through existing literature and domain interviews, we attempt to build a more comprehensive list of techniques (\textbf{DC1}). We balanced the technical feasibility and the domain significance in our list.

\textbf{Eye contact} is a technique we found important enough to focus on.
DE2 claimed that in online contests ``\textit{connection to the audience with a camera is critical. The key is to be a natural speaker in front of the camera.}'' DE1 and DA1 also found its importance, with DA1 claiming they ``\textit{don't know where to look at.}''. Literature on online public speaking also emphasized the importance of eye contact. In the book \textit{Presenting Virtually} \cite{presentingvirtually2021}, Patti Sanchez claimed ``eye contact is essential to make your message feel direct and personal'', however, ``looking directly into the camera can be a challenge for presenters who are accustomed to speaking in person''. 
Data from online speeches provides the ability to assess the effective use of eye gaze, since there is a universal viewing angle of the speech by all online members.

\textbf{Body gesture and use of stage} are important for online speeches.
In open discussion, four of the six interviewees thought the impact of the screen in online interactions for body language was especially important. 
However, some of them also mentioned the difficulty of keeping various motions within the limited space of screen. This also reflects the importance of the proper use of ``stage'' in the camera. DE2 questioned: ``\textit{You are the director with a fixed camera. Do you need to show your full body above your waist?}''
Online speaking enables accurate calculation of stage and gesture occlusion by providing a consistent camera angle, unlike offline speech where the audience's viewing angles vary.

\textbf{Emotion} plays an important role in public speaking. In our interview survey, emotion related factors ranked high in the list while emotional diversity was ranked the most important factor by interviewees. Emotion can be explicitly expressed by the speakers' facial expression, voice and text content. In our review of several methods of extracting vocal and textual emotion, we found them less accurate than the facial expression results, even with the state-of-the-art models. Therefore, this work mainly focuses on facial expressions.

\textbf{Vocal characteristics} include voice volume, pitch, speaking rate, and pauses. Proper speaking rate and pauses are difficult because there is no feedback from the audience in the online contest. DA1 stated ``\textit{Faces of the audience are not shown, and no audio reply is allowed. You don’t know if the audience gets your points or joke, and can’t change without feedback.}'' 
While volume and pitch were not mentioned by the interviewees, they are commonly considered important in public speaking literature. Online speakers are required to have good volume management skills to control their speaking volume.

\section{System and Data}

\subsection{System Overview}

\begin{figure}[tb]
 \setlength{\belowcaptionskip}{-0.6cm}
 \setlength{\abovecaptionskip}{0.2cm} 
 \centering 
 \includegraphics[width=\columnwidth]{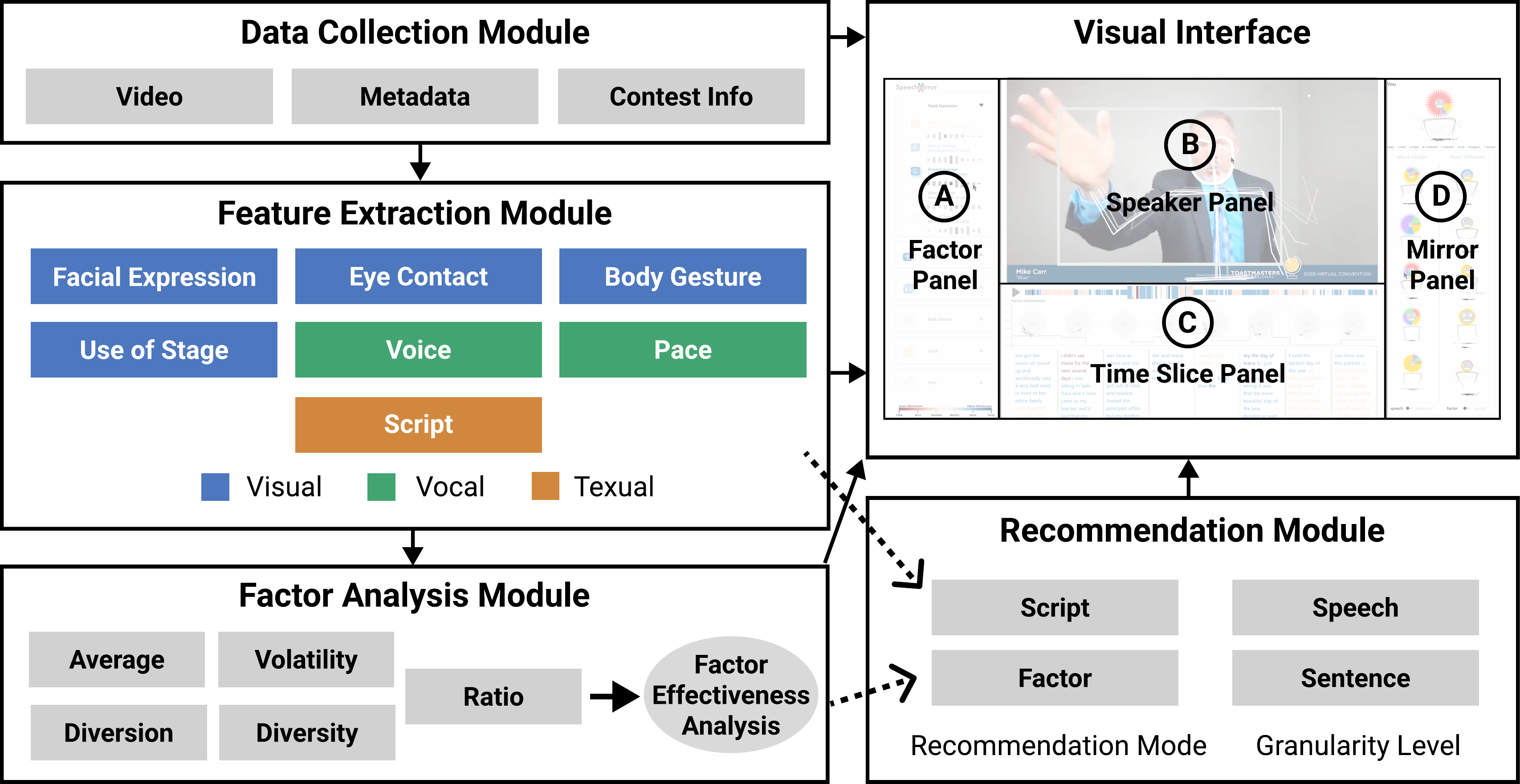}
 \caption{System architecture of \textit{SpeechMirror}.}
 \label{fig:architecture}
\end{figure}

We design and implement a visual analytic system, \textit{SpeechMirror}, to fulfill the analytical goal and the design considerations. 
Our system integrates raw video data, extracted multimodal feature data, speech factors and corresponding effectiveness data to provide a complete data processing and analyzing workflow. All data except the raw video is stored in MongoDB for fast access.

As illustrated in \autoref{fig:architecture}, our system consists of five major modules. 
The \textit{Data Collection Module} contains the speech videos and information we collected for analysis. The newly input video will also be stored in the module.
The \textit{Feature Extraction Module} extracts features from the input video, including visual, vocal and textual modalities.
The \textit{Factor Analysis Module} determines speech factors based on the extracted features and further estimates the effectiveness of factors.
The \textit{Recommendation Module} searches for the most similar and different speeches from the video collection.
The \textit{Visual Interface} provides visualizations of data and interactions to support analysis. The interface is designed and implemented in a browser-server architecture, utilizing d3.js \cite{Bostock2011d3} for creating visualizations in the front-end interface, and Flask framework \cite{flask} for providing web services on the back-end.

\subsection{Data Collection Module}

We manually collected a collection of speech videos in the World Championship of Public Speaking (WCPS) contest from public online platforms like YouTube.
The entire online speech video collection in our system contains 102 videos in total. Each speech is about 7 minutes long and of good visual-audio quality. 
We recorded the metadata (including the start and end of each speech in the video), and contest information (including region, year, level, and rank). Whether the video is delivered online or offline was also labeled.

\subsection{Feature Extraction Module}

To assist users with quantitative analysis of presentation techniques, the feature extraction module takes video as input and automatically extracts and processes features for further analysis.
The module extracts the following speech-related features from multimodal inputs including visual, vocal, and textual.
The multimodal features are aligned based on the timestamps of the script's words.

\begin{itemize}[leftmargin=*]
\vspace{-0.2cm}
\setlength{\itemsep}{0pt}
\setlength{\parsep}{0pt}
\setlength{\parskip}{0pt}

\item \textbf{Facial expression:} The face of speaker in each frame is detected by face\_recognition \cite{face_recognition} and clustered out from other faces by DBSCAN \cite{dbscan}. We apply AffectNet \cite{affectnet} to predict valence and arousal values (ranging from [-1, 1]) from face images. AffectNet is a widely used baseline method for predicting facial expression, valence and arousal from images in the wild. A convolutional neural network \cite{emotiontype} is used to classify face images into seven classic emotion categories \cite{fer2013}.

\item \textbf{Eye contact:} We apply OpenFace Toolkit \cite{openface, eyegaze2015} to estimate the eye gaze direction of both eyes. The eye gaze direction of each eye is represented as a normalized 3D vector in world coordinates. Average eye gaze direction is converted to radians in world coordinates. 
The angle of watching the camera is calculated by averaging the angle obtained from the 3D position vector of the eyes relative to the camera and the vector of the eye gaze direction.

\item \textbf{Body gesture:} We adopt MMPose \cite{mmpose2022}, a widely used open-source toolbox for pose estimation, to predict the 2D body keypoints from videos. 
We use Faster R-CNN model \cite{fasterrcnn} with a ResNet-50-FPN backbone for human bounding box detection. 
HRNet \cite{hrnet} pre-trained on COCO \cite{coco2014} detects 2D keypoints from video frames with the Human3.6M \cite{human36m} format.
We set up rules to retain the body keypoints of the speaker. 
We calculate the kinetic energy \cite{niewiadomski2013human} of the speaker's upper body as the gesture energy based on the offset of the keypoints in adjacent frames and the mass distribution of human body \cite{plagenhoef1983anatomical}. 
Gesture diversity is determined by the standard deviation of the cosine distances between the upper body keypoints of each frame and the first frame. The body keypoints are first aligned by the thorax and normalized to a fixed width of shoulder.

\item \textbf{Use of stage:} We utilize the $z$-axis position of the speaker's head relative to the camera estimated by OpenFace Toolkit \cite{openface}, as the distance between the speaker and the camera. The position of the speaker in the video frame is determined by the center of the bounding box detected by MMPose during body gesture extraction.

\item \textbf{Voice:} The loudness and pitch data is extracted by Parselmouth \cite{parselmouth}, a Python library for Praat \cite{praat} which is a widely used speech analysis software in phonetics. 
We use the ``sound to intensity'' and ``sound to pitch'' functions to capture the loudness (in dB) and frequency (in Hz) of a speech, respectively.

\item \textbf{Pace:} We compute the pauses between different words and different sentences according to the timestamps of each sentence and word. The speaking rate is determined by the duration per syllable of each word. The syllable of word is counted by the vowel sounds in a word with the CMUdict corpus \cite{cmudict} in NLTK Language Toolkit \cite{bird2009nltk}.

\item \textbf{Script:} Each video is transcribed by Azure Cognitive Speech to Text Service \cite{azurespeech2text} with timestamps of each sentence and word. We use the Universal Sentence Encoder \cite{cer2018universal} to encode script texts into 512 dimensional vectors while maintaining the semantic information.

\end{itemize}

\subsection{Factor Analysis Module}

The visual and vocal features extracted from the videos are time series, which are difficult for understanding and comparison. Considering the domain requirements, we calculate the factors as shown in \autoref{tab:pretech}. 

The methods of factor calculation are as follows: 
\textbf{Average} represents the mean value of data over time.
\textbf{Volatility}, representing data change over time, is calculated using the CID algorithm \cite{batista2014cid} with normalization. CID measures the complexity of time-series data, capturing patterns like peaks and valleys.
\textbf{Dispersion} is determined by the variance of the time-series data, obtained by dividing the standard deviation by the mean.
\textbf{Ratio of watching camera} indicates the proportion of frames that the speaker is looking directly at the camera within a 5-degree angle.
\textbf{Diversity of facial expression type} represents the variety and relative abundance of the emotions \cite{quoidbach2014emodiversity}. The time-series emotion type data are represented as $D=\{d_t\}_{t=1}^T$, where $d_t$ indicates the $t$-th sample of emotion type data. 
Let $r_i$ denote the proportion of the same emotion type as $d_i$ in $D$. Diversity is calculated in  $diversity = \sum_{i=1}^{T}(r_i\times \ln{r_i})$.

With the speech factors and contest placements of speeches indicating speech effectiveness, we build the effectiveness relationship of each factor with a regression method \cite{gutierrez2015ordinal}. Specifically, we conducted parallel line tests and observed significance with $p<0.05$. Then a multi-class ordinal regression is employed to evaluate the significance of speech factor effectiveness and the regression results are presented in \autoref{tab:pretech}. We predict the effectiveness of each factor based on its significant relationship with speech effectiveness.

\subsection{Recommendation Module}

To enable users to quickly find the presentation examples that are similar or different to the speech for analysis (\textbf{DC2}), our system recommends relevant speeches from our data collection for their reference.

The recommendation module grants users the option to manually select from \textbf{two granularity levels} (speech or sentence) and \textbf{two recommendation modes} (factor or script) through the interface. The granularity levels determine the extent of the search range, which spans either the entire speeches or individual sentences. The recommendation modes dictate the method of similarity calculation, which can either rely on speech factors or the transcribed script content.

The recommendation consists of three steps: (1) Prepare the query and candidate data. The module calculates the query data of selected period in the analyzed speech, as well as the data of candidates according to the selected granularity level. (2) Extract the feature vectors for both query and candidates. For the factor recommendation mode, we join the values of the factors selected on the interface into a vector after a min-max normalization. 
For the script recommendation mode, we use the textual semantic embedding vectors for both input and candidate scripts.
(3) Fetch the most similar or different candidates to the query. We calculate the similarity distances between the query and the candidates by Euclidean distance for vectors in the factor mode and cosine distance for vectors in the script mode. Heap queue algorithm is used to obtain the results with largest or smallest similarity.

\section{User Interface Design}

In this section, we introduce the general principles for design, the interface, and visualization designs.

The design of our visualizations aimed to be intuitively understood by an audience with minimal visualization literacy, which would allow our interface to be used by a wider audience with minimal training. 
We iterated our design closely with potential users and in the process created design principles that guided our work. In our initial designs we found significant difficulty in understanding the concepts in our interface. In order to increase the intuitiveness of our system, we sought design principles that followed several principles given by Blair et al. \cite{blair2008user} and reflections by Böttinger et al. \cite{bottinger2020reflections}, including: \textbf{(DP1)} making consistent use of visual elements, \textbf{(DP2)} providing elements in proximity to the content, \textbf{(DP3)} offering interactions as direct with content as possible, \textbf{(DP4)} showing understandable explanatory visualizations to a broader audience. The design alternatives in our design iteration process are introduced in Section. 4.1 of the supplemental material.

\subsection{Icons and Color Encodings}
\label{sect: icon-color-encoding}

\begin{figure}[tb]
 \setlength{\belowcaptionskip}{-0.6cm}
  \setlength{\abovecaptionskip}{0.2cm} 

 \centering 
 \includegraphics[width=\columnwidth]{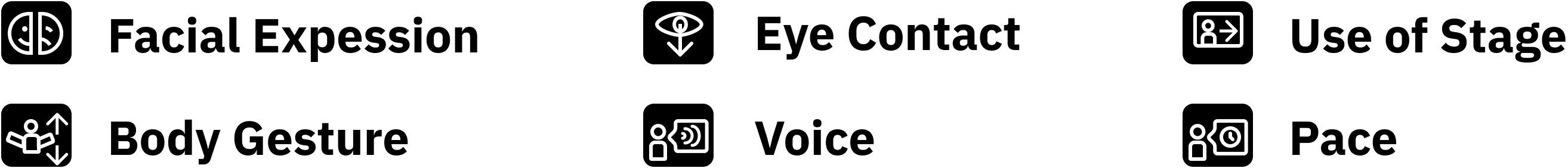}
 \caption{Consistently used icons for the presentation techniques.}
 \label{fig:color-icons}
\end{figure}

Repetition of elements in the design of icons, as shown in \autoref{fig:color-icons}, was used to both aid in the understanding of intricate concepts and used throughout the three panels in the system to allow rapid understanding. \textbf{(DP1)} The icons were designed as pictographs that resemble the concepts they are linked to. For example, the position of a speaker on the screen was given by a figure icon surrounded by a screen. 

We applied consistent use of color encoding for effectiveness across the three panels of our interface using a diverging color scale that emphasizes the data of two extremes, as shown in \autoref{fig:teaser} (A1). \textbf{(DP1)} This color scheme is a scale with dark red signifying very low effectiveness metrics to a dark blue signifying very high effectiveness metrics. A light gray is used to indicate factors that were not significantly related to effectiveness. 
Another color encoding scheme for emotion is employed in SpeechTwin, which will be further introduced in \autoref{sect:speechtwin}.

\subsection{Interface Panels}

Our interface is composed of four panels that support the design considerations as described in \autoref{sect:design-considerations}.

\textbf{The Factor Panel} presents an overall view of factors that appear in a speech, 
and how they relate to all speeches in our video collection (\textbf{DC1}), as shown in \autoref{fig:teaser}-A. 
The 23 speech factors considered in our study are categorized into 6 groups of presentation techniques. 
The panel provides a summary of the factors when a group is closed, and more detailed information when opened.
Each factor is represented by colored icons and text that indicate its effectiveness, as explained in \autoref{sect: icon-color-encoding}. \textbf{(DP4)}
When a user hovers over a factor, the factor effectiveness board (A2) will be displayed, showing the effectiveness trend and the factor distribution in the video collection. 
Factors selected in this panel will filter the results in the other three panels. If no factor is selected, then the aggregated results of all factors are shown. 

\textbf{The Speaker Panel} supports understanding speaking techniques with direct reference to the original context in the video. The techniques are visually presented, overlaid on the video with reference to the regions they correspond to. Users can directly interact with the layered visualizations to focus on a specific technique, thereby displaying the presentation technique board (B1). 
Data shown in the panel comes from the current playing sentence. Once a factor is focused, more detailed information about the use of the factor is shown, as well as a  recommendation of speeches for reference
(\textbf{DC2}). 
Users are allowed to play the recommended videos and set one to the comparison video. The comparison video will be displayed at the top-right corner of the Speaker Panel and can be switched to the main video in the panel.

\textbf{The Time Slice Panel} enables users to select different parts of a speech, as well as provides different views of speech data with script context over time (\textbf{DC3, DC4}), as shown in \autoref{fig:teaser}-C.
At the top of the panel, the timeline (C1) shows the use of selected factors over time. Each rectangle represents a sentence, with the color encoded as described in \autoref{sect: icon-color-encoding}. White blanks indicate the intervals between sentences. A brush tool is provided for convenient selection of speech periods on the timeline. 
The selected period is divided into 8 time slices, using glyphs (C2) to show more detailed use of the factor and raw data. At the panel bottom is a text module (C3) highlighting the effectiveness of factors in reference to the speech script.
Words that coincide with the time slice segmentation points are split into two parts based on time, allowing for a better comprehension of the speaking pace within shorter selected time periods.

\textbf{The Mirror Panel} allows speakers to find the most similar and most different speeches compared to the whole or a part of a speech (\textbf{DC2}), as shown in \autoref{fig:teaser}-D. 
At the top of the panel, a visual summary of speech factors (D1) for the selected part of the speech, named SpeechTwin, is presented to facilitate quick comprehension of speech factors.
Below, the panel demonstrates the SpeechTwins of similar and different speeches (D2) according to the selected recommendation mode and granularity level (D3). Hovering over a SpeechTwin will trigger the speech comparison board (D4) while clicking on it will set the speech as the comparison video. Users are allowed to switch between the original video and the comparison video by clicking the video name located at the panel top.

\subsection{Visual Design}

\subsubsection{SpeechTwin: A Multimodal Speech Summary}
\label{sect:speechtwin}

\begin{figure}[tb]
 \setlength{\belowcaptionskip}{-0.6cm}
 \setlength{\abovecaptionskip}{0.2cm} 

 \centering 
 \includegraphics[width=\columnwidth]{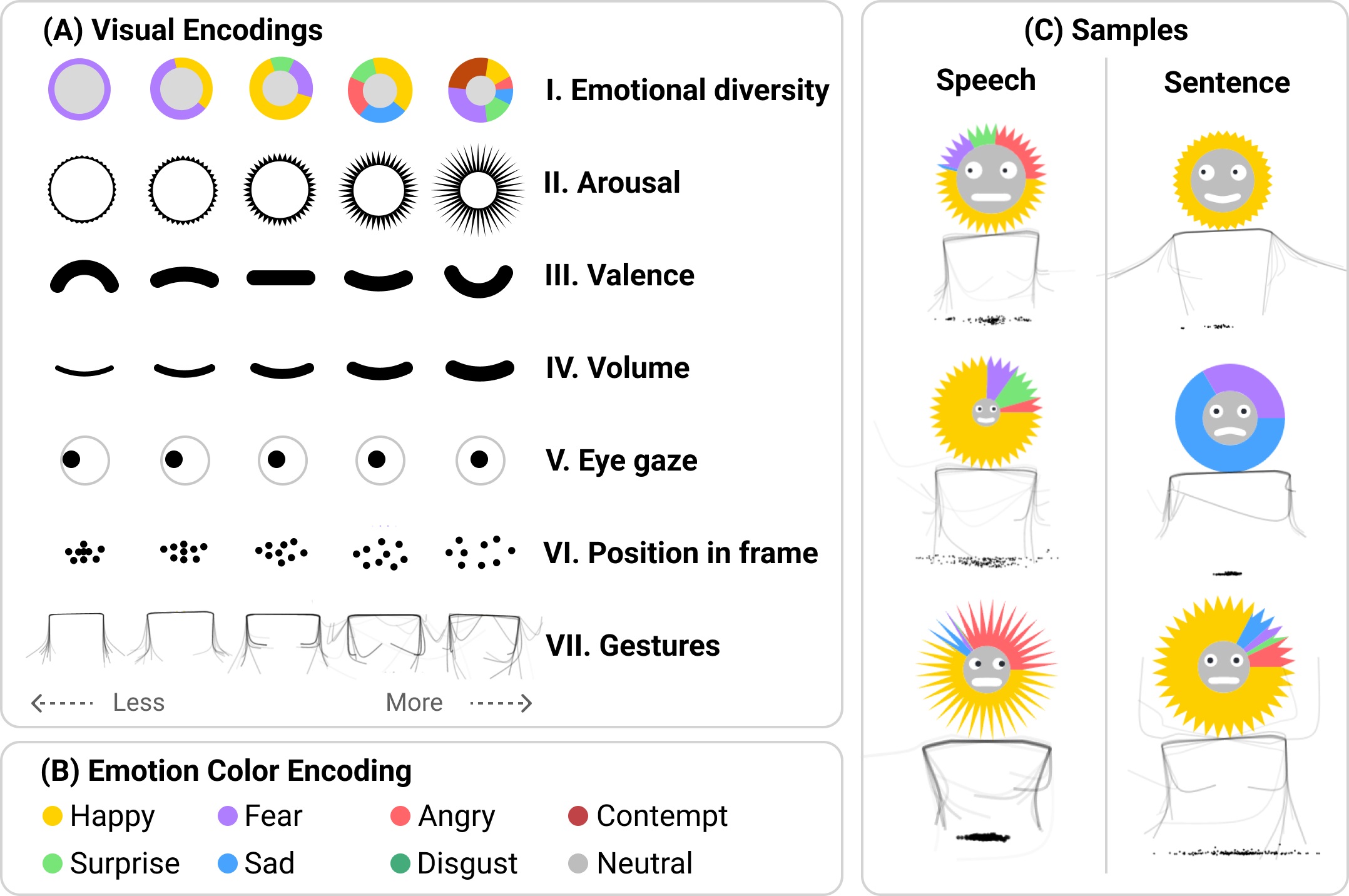}
 \caption{SpeechTwin: a novel visualization of multimodal speech summary.}
 \label{fig:speechtwin-demo}
\end{figure}

Our system provides a visual summary of crucial speech techniques in an intuitive way by 
mapping the technique to symbols on a figure  (\textbf{DC5, DP2}).
We display the visual encoding used and the demonstration of SpeechTwin in \autoref{fig:speechtwin-demo}. More demonstrations are shown in the supplemental material (Sect. 4.2) and supplemental video for reference.

Facial and vocal data are represented on a Chernoff face \cite{chernoff1973use}, which allow these features to be combined and closely associated with corresponding facial regions. 
Eye gaze direction is communicated by the angle the eyes are looking at (\autoref{fig:speechtwin-demo} A-\uppercase\expandafter{\romannumeral5}). 
The positivity or negativity of facial expression, namely valence, is given by the upward or downward turn of the mouth as a smile or frown (A-\uppercase\expandafter{\romannumeral3}). 
The intensity of facial expression, arousal, is conveyed by the protrusion of spikes outside the face, as inspired by the coding of arousal in shape \cite{sievers2019multisensory} (A-\uppercase\expandafter{\romannumeral2}). 
The emotional diversity is indicated by the area of face given to emotions (A-\uppercase\expandafter{\romannumeral1}), with a neutral gray emotion in the center, and other emotions shown in the area of the outside circle and with different colors displaying different emotions (\autoref{fig:speechtwin-demo} B).
The loudness of voice is represented by the width of the mouth (A-\uppercase\expandafter{\romannumeral4}). The eyes and the mouth maintain a fixed proportion in relation to the face. To ensure the readability of the elements inside the face in rare cases of scarce neutral emotion, we set a minimum size to the face.

To better describe various gestures in the speech while minimizing visual distractions, the representative body gestures are given in the arms and shoulders of SpeechTwin. The upper body keypoints of the speaker in each frame are aligned by the position of the thorax, and normalized to a fixed shoulder width. Inspired by PoseTrans \cite{jiang2022posetrans}, we use a Gaussian Mixture Model (GMM) to classify the poses into 10 clusters. We identify the pose with the smallest sum of cosine distances to all other poses in the cluster as the most representative one. We visualize the most representative gestures with the opacity indicating the amount of gestures in the corresponding cluster (A-\uppercase\expandafter{\romannumeral7}).

The use of the stage is described by the ``footprints'' or dots beneath the character, with each dot representing the center of the speaker's bounding box on the screen (A-\uppercase\expandafter{\romannumeral6}).

Hovering over a SpeechTwin in the Mirror Panel (\autoref{fig:teaser}-D2) will trigger the speech comparison board (D4). On the board all factors of the compared speaker are contrasted with the original speaker. Factor differences between the speeches are shown on a bar chart, with the higher differences polarized at the top and the bottom. The placement of the speech in the contest is also shown. 

\subsubsection{SpeechPlayer: An Augmented Speech Video Player}

Considering the potential information omission when viewing speech videos, we offer SpeechPlayer (\autoref{fig:teaser}-B), an interactive approach to enhancing users' understanding of techniques during video playback, as well as augmenting their sense of expressive possibilities (\textbf{DC4}).

SpeechPlayer integrates the visualization elements of critical presentation techniques directly in the video feed. \textbf{(DP2, DP3)} The facemark keypoints are displayed on the video frame to emphasize the facial expressions of the speaker (\autoref{fig:teaser}-B4). The direction of eye gaze is depicted through a ray, which becomes increasingly transparent along the direction of the gaze (B3). 
We found with traditional methods of an opaque line, confusion about the direction of eye gaze can occur when viewing eye gaze over time, such as when speakers change their head angle to act different characters.
The skeleton of the speaker's upper body (B5) and the bounding box of the speaker (B2) are also visualized to enhance the understanding of body gestures and the positions in video frames.

To enhance the understanding of eye gaze and body gestures across time, SpeechPlayer displays 10 skeletons and eye gaze rays at fixed time intervals. 
The transparency of color corresponds to the temporal proximity to the current playback time, with higher opacity indicating closer proximity to the current time. 

Users can interact directly with the displayed elements on SpeechPlayer, and hovering allows analysis of a group of presentation techniques with the presentation technique 
board (\autoref{fig:teaser}-B1). Users can choose a speech factor of interest, and compare the distribution of the factor values of the currently played sentence with the average factor value of best speeches and all speeches (\textbf{DC1}). Speeches or sentences that have the most similar and most different use of a factor are recommended so users can rapidly find reference samples (\textbf{DC2}).

\subsubsection{Visualizations of Multimodal Features}

\begin{figure}[tb]
 \setlength{\belowcaptionskip}{-0.6cm}
 \setlength{\abovecaptionskip}{0.2cm} 

 \centering 
 \includegraphics[width=\columnwidth]{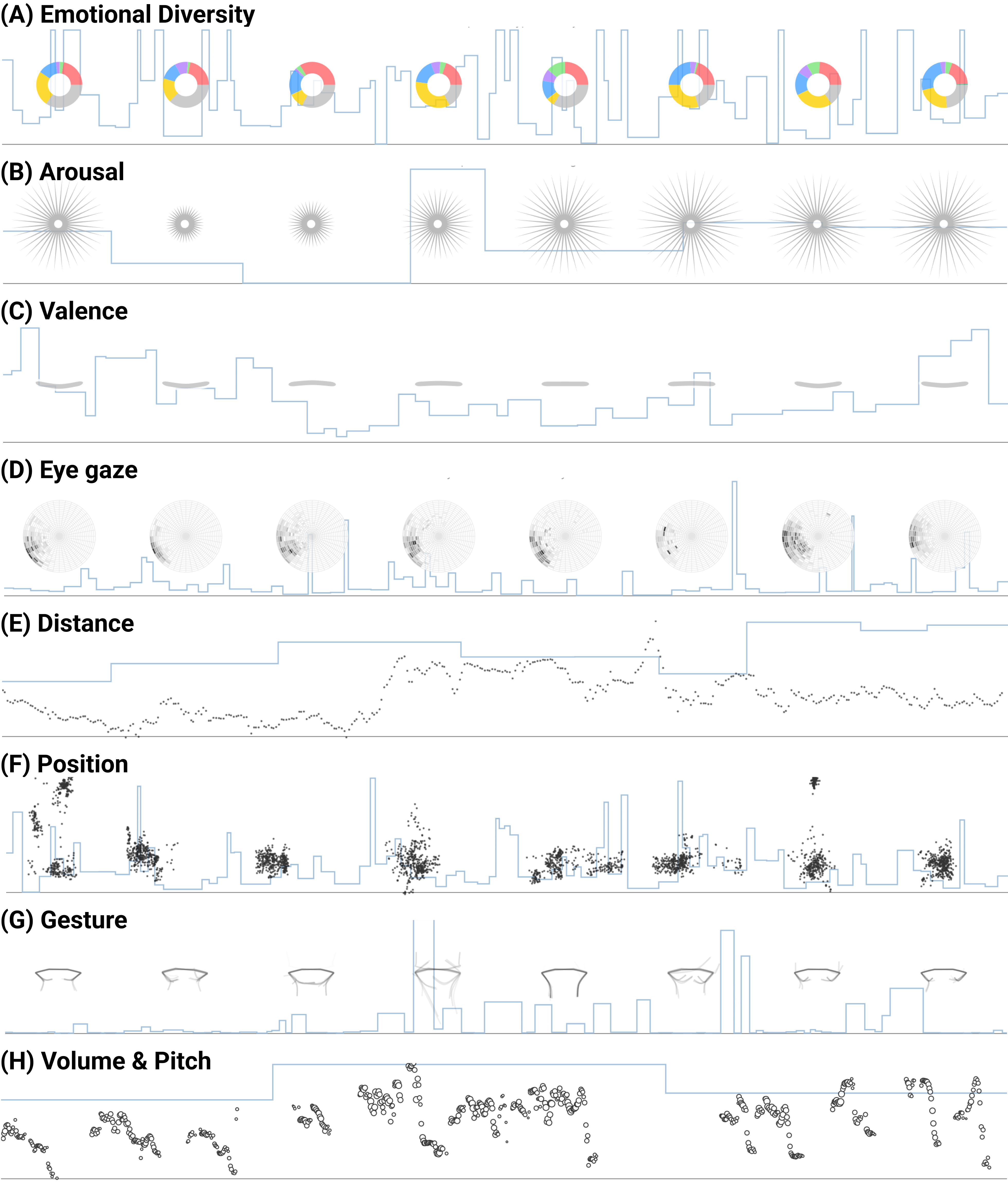}
 \caption{Demonstration of modality feature visualizations.}
 \label{fig:modalityfeature-demo}
\end{figure}

We offer visualizations in the Time Slice Panel to illustrate the temporal details of modality features. These visualizations are tailored to the specific nature of original feature, as shown in \autoref{fig:modalityfeature-demo}.

For each feature in facial expressions, we use visualization methods consistent with SpeechTwin, as shown in \autoref{fig:modalityfeature-demo}-A, B and C respectively. \textbf{(DP1)}
In an informative visual summary of gaze patterns, a gaze heatmap is used to depict the frequency of eye gaze directions. Darker shades on the heatmap indicate higher frequencies.
To better understand body gestures, we considered several visualization methods as mentioned in \autoref{sec:human-physical-behavior}. Balancing between the need for simplicity and complexity of gestures in a large number of video frames, we display representative gestures that provide an uncluttered view of representative gestures, as shown in \autoref{fig:modalityfeature-demo}-G.
The data in each time slice are aggregated to summarize the features.

We offer two visualizations to depict the use of stage techniques. 
To understand the speaker's positions in the video frame, we display the centers of speaker's bounding boxes in each time slice within a rectangular area, as shown in \autoref{fig:modalityfeature-demo}-F. This allows us to observe the range of speaker's position changes throughout the time slice. 
Additionally, we apply a scatter plot to illustrate the changes in distance between the speaker and camera over time in the selected speech period.

For the features in the voice technique, we use a scatter plot to visualize volume and pitch data along time referring to the work of Schaefer et al. \cite{schaefer2016intuitive}, as illustrated in \autoref{fig:modalityfeature-demo}-H. We map the $x$-axis as the timeline, with the $y$-axis representing pitch, and the radius of each point indicating volume. Users can observe the changes in volume and pitch over time directly through this chart.

\subsection{Interactions}

The proposed system, \textit{SpeechMirror}, enhances users' analysis abilities while maintaining intuitiveness and fluency through various ways of interaction within the four linked panels of the system. \textbf{(DP3)} Here we summarize the supported interactions:

\textbf{Clicking.} 
In the Time Slice Panel, clicking a word within the text module will trigger the video in the Speaker Panel to jump to the corresponding start time of the word. Double-clicking the word will select the sentence that contains the word on the timeline.
Users are allowed to activate the comparison video in the Speaker Panel by clicking the SpeechTwin in the Mirror Panel. Clicking on the video will play or pause the comparison video, and double-clicking will switch the video in Speaker Panel to the comparison speech, along with changing the corresponding visualizations in the Time Slice Panel.

\textbf{Selecting.} Our system enables users to select any factor or a group of factors in a feature for analysis in the Factor Panel. The selection of a factor will result in the Time Slice Panel switching to the corresponding factor data, and the SpeechTwin of the analyzed video in the Mirror Panel being refreshed, thus generating new recommendations.
Users are allowed to select different recommendation modes and different granularity levels with the toggle buttons in the Mirror Panel.  

\textbf{Hovering.} 
Hovering on the histogram in the Factor Panel displays the factor effectiveness board, while hovering over the data view in the Speaker Panel reveals the presentation technique board for more detailed information.
In the Time Slice Panel, when users hover over the line chart of factor distribution, the time-aggregated data of the hovered feature will be shown above. Furthermore, when users hover over the SpeechTwin in the Mirror Panel, a new panel will be shown to reveal the difference between factor values of the speakers.

\textbf{Dragging and Brushing.} In the Time Slice Panel, users can drag the white triangle on the timeline to change the play position on the video. Users can brush a selection area on the timeline or drag the selected area to focus on a sequence of sentences. The Time Slice Panel and the Mirror Panel will refresh accordingly.

\section{Evaluation}

We aim to evaluate the proposed system and visualizations in two scenarios based on the framework introduced by Lam et al. \cite{lam2011empirical}: user experience (UE) and visual data analysis and reasoning (VDAR). 
We used a user experience questionnaire and a follow-up interview to elicit the subjective feedback and opinions on our system and visualizations (UE). A case study is used to demonstrate how users explore data and find insights with our system (VDAR).

\subsection{Study Design and Procedure}

A user study was designed to assess the performance of our system and visualizations in the two scenarios. 

Since experts and amateurs in public speaking are the target users of our system, we aimed to gain insight into their diverse usage patterns and perceptions of the system. For amateur participants, we requested their recent online public speaking videos for analysis in the evaluation. Expert participants were asked to analyze a speech from our data collection to simulate critiquing a student's speech as a coach.

The procedure of our evaluation study was split into three sessions. The whole study lasted about 90-100 minutes.

\textbf{Introduction.} We first introduced the main goal and background of our system. After obtaining informed consent from the participants and collecting their personal information, we provided a detailed introduction of the important concepts to facilitate their understanding and utilization during subsequent sessions. By means of quick questioning, we confirmed that participants had fully understood the concepts we presented. The introduction session took about 20 minutes.

\textbf{Exploration.} The user study contains two phases for the participants to explore our system: user tasks and free analysis. 
We guided the users to complete the user tasks while we observed their use of our system and assessed usability.
The free analysis phase aims to observe how participants analyze speeches with our system by allowing them to use the system on their own. They were encouraged to think aloud while using the system and voice their findings during the analysis.
Both phases took about 30 minutes.

\textbf{Reflection.} Based on their experience with the system, participants were requested to complete a user experience questionnaire, and were encouraged to provide their opinions for each question. We further conducted a follow-up interview to gather additional feedback and insights about our system and its potential actual applications. This session took about 10 minutes.

\subsection{Participants}

We recruited 8 participants for the evaluation: four experts (EE1 - EE4) and four amateurs (EA1 - EA4) in public speaking. None of the participants participated in the previous domain interviews. 
The experts all have more than 3 years in coaching for public speaking and have participated in the contest. The amateurs did not coach public speaking as an occupation, and all had participated in the contest. 
The participants were compensated for \$94 per expert user and \$54 per amateur user. Detailed information of the participants in the user study is provided in the Sect. 5.1 of the supplemental material.

\subsection{User Experience Questionnaire}

\begin{figure}[tb]
 \setlength{\belowcaptionskip}{-0.6cm}
 \setlength{\abovecaptionskip}{0.2cm}

 \centering 
 \includegraphics[width=\columnwidth]{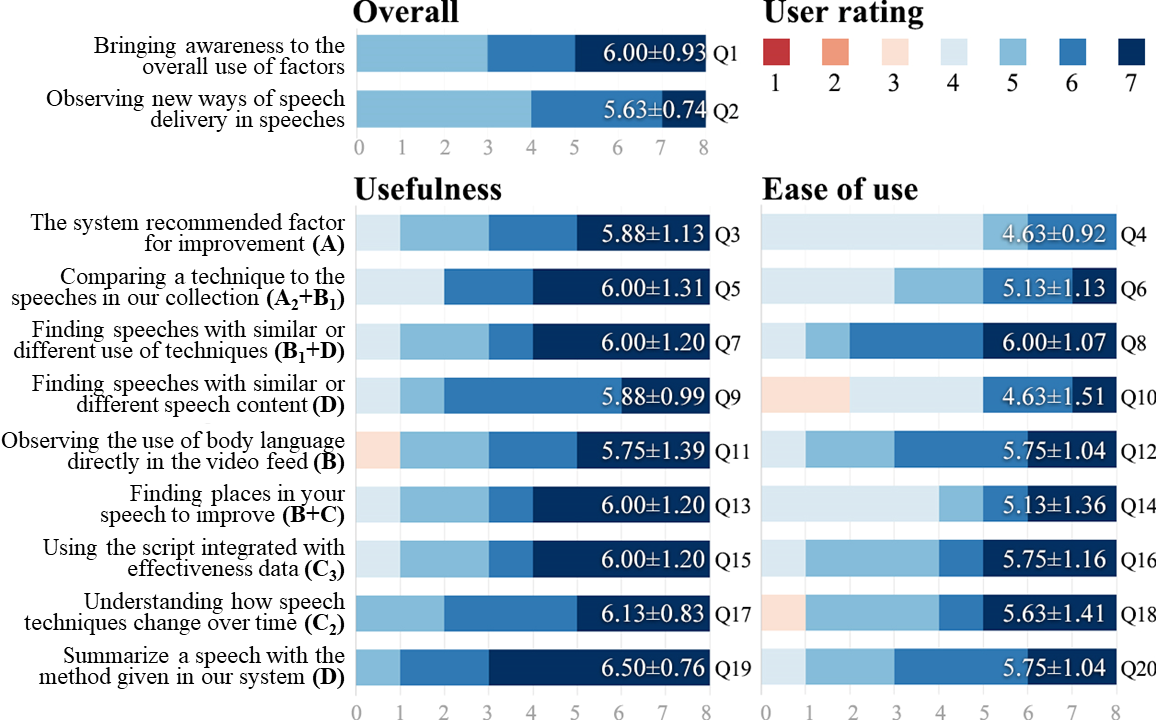}
 \caption{Questions and results of our user experience questionnaire. The white overlay in the bars indicates the Mean±SD. The target of each question is indicated by the labels in \autoref{fig:teaser}.}
 \label{fig:evaluation-userexperience}
\end{figure}

A user experience questionnaire was employed to assess the participants' subjective feedback of our system and visualization approaches (UE). 
We adopted a 7-point Likert scale to score the experiences of participants. The questions with their evaluation target and corresponding scores are illustrated in \autoref{fig:evaluation-userexperience}. We present the ratings of all participants for each question in the supplemental material.

In general, participants appreciated that our system helped them become aware of the usage of presentation techniques in speeches (Q1) and discover new ways of expression (Q2). 
On both of these questions, our system received an average score of 5.81 out of 7, which indicates the participants' overall acceptance of \textit{SpeechMirror}.
For the usefulness and user-friendliness of use of different parts of our system (Q3-Q20), we received an average score of 5.69, indicating the participants' general approval of the methods we adopted. 

The participants acknowledged the usefulness of our system, with an average score of 6.01/7 for usability. Specifically, the participants were satisfied with the usefulness of viewing techniques over time (Q17 for the Time Slice Panel and the visualizations of modality features) and the summary of presentation techniques (Q19 for the SpeechTwin).
One amateur participant rated 3/7 on the usefulness of the Speaker Panel and SpeechPlayer (Q11), while other participants provided positive ratings. 
She suggested providing an option to enhance the original video with an on/off toggle or in a separate window.

The participants' satisfaction with the ease of use was lower, with an average score of 5.38. 
There are two questions in our questionnaire that were rated below 5.0 on average.
The concept of factor effectiveness and related charts required more effort to understand for participants, especially those without a STEM background. Additionally, providing users with multiple factors could also result in cognitive burden, as users reported feeling unsure about where to start exploring factors. (Q4 for the Factor Panel)
Recommendation based on speech content was rated as 4.63. Several participants found the recommendation results inaccurate. This is likely due to our limited database and lack of samples for comparison. (Q10 for the Mirror Panel)

\subsection{Follow-up Interview}

We conducted a semi-structured interview to gather further feedback and insights on \textit{SpeechMirror} from the participants (UE). 

\textbf{Factor Analysis.} Our system received positive feedback on providing nearly comprehensive speech effectiveness analysis. EA3 said ``\textit{It is amazing to be aware of so many factors. I didn't imagine this was possible.}''
EE3 commented that ``\textit{all these factors are very useful}''.
EE4 considered the analysis of our system to be very interesting. After watching the whole speech of his ``student'', he confirmed that the result of our system matched his judgment of the speech as an expert coach.
However, EA1 mentioned feeling information overload, feeling ``\textit{overwhelmed due to a lot of factors}''. 

\textbf{Speech Recommendation.} The participants appreciated the recommendation function since it took less effort to find other speeches for comparison. 
EA2 said that ``\textit{I have never thought about who is the most similar or different to myself. Before, it was hard for me to find someone to compare to.}''
EE1 and EE2 both mentioned that the SpeechTwin demonstrated the ``speech style'', allowing users to compare it with others and ``\textit{find a personalized speech style}''.
EE2 and EA1 stated the benefit of the Mirror Panel and SpeechTwin in enabling users to ``\textit{find a comparison and reference}'' and ``\textit{know who to learn from}''.

\textbf{System Complexity.} The participants indicated a preference for a simplified system to improve user experience, suggesting that the current system is overly complex. 
EE1 reported a high learning curve and instead wished for more ``\textit{visual reminders for key functions and terms.}'' 
EA2 and EA3, with non-STEM backgrounds, also suggested adding tooltips in the system for better understanding.
EE4 suggested a potential simplification of the interface by ``\textit{filtering the most important elements}''. 
EA2 and EE4 observed that the text script within the Time Slice Panel was not very legible, potentially because there were too few words on each line in each slice.

\textbf{Potential Application.} In the interview, we gathered valuable insights from the participants regarding the potential applications of our system. We also observed a notable level of anticipation for further utilization of the system.
EE3 stated that she planned to have students review the system-generated results and analyze presentation techniques with the system in tandem with her class material.
EA3 noted ``\textit{I want to put all my speech videos into the system and see where to improve. The system directly tells me how differently others present, so I can learn from others}''. We see a promising application in the future of our system or other similar systems.

\subsection{Case Study}

During the free exploration phase of our evaluation, both experts and amateurs found unexpected uses of \textit{SpeechMirror} (VDAR).
We provide more details in our supplemental document (Section 5.5) and video.

\subsubsection{Expert user: Coaching with SpeechMirror}

EE1, a district champion for the Toastmasters speech evaluation contest and a university and youth speech coach noticed from the Factor Panel that the speaker he was analyzing had exceptionally low arousal, most similar to speakers at the club level. He then used the Mirror Panel to find one of the most different SpeechTwins. He noted the speaker was the previous year's world 2nd place winner. He then switched the video to the comparison video and used the timeline to find a moment of especially high arousal. After noticing the changes in arousal over time, he said ``\textit{I think the tool could be very useful. In order to build a connection to the audience, you need contrast, you need the twists and turns of emotions. Then you can elicit emotions of the audience. I can use this to show my student that there is strong contrast [for this expert speaker]. I could let the student see where they are and compare with different speakers.}'' While EE1's mouse was hovering on a SpeechTwin with a smooth spread of positional footprints, EE1 added, ``\textit{With the SpeechTwin we can see where they are and compare with the other speakers to see possibilities.}'' 

He then moved his mouse to indicate a SpeechTwin with a smooth distribution of footprints and commented ``\textit{this speaker had constant stage movement.}'' He then pointed at a SpeechTwin with a clear three part distribution of footprints,  
stating ``\textit{if like this there is a clear stage design}''. He noted the possible use of our system for coaching, ``\textit{Overall I feel like this is a tool that I could use and discuss with my students and find out different insights so that they can improve.}''

\subsubsection{Amateur user: Improving with SpeechMirror}

EA2 was a previous district contestant that had attended 2-5 online contests. During evaluation, she used our system to better observe changes in her valence and to find more possibilities of improving her gestures. She first noted that while overall, the average valence in her face was indicated by the system as performing well, the system found her speech most similar to someone that was smiling throughout their speech. She guessed it might be that her sadness wasn't expressed obviously enough. When looking at the Time Slice panel, she observed that the ending of the speech was happy, ``\textit{This is correct.}'' However, she then pointed to earlier time slices in her speech ``\textit{For these two parts I really wanted to show my sadness, my anger, but I guess I failed.}''

While comparing by selecting both the valence average and gesture energy change, she stated ``\textit{I want to see who is the most similar to me.}'' She selected the most similar speech, noting that the gestures, similar to hers, were mostly below the screen or on the screen border.

She then looked at the Mirror Panel to find the most different speaker. When she played the video, she saw a speaker that had many different gestures and stage changes.
While looking at the speech comparison board to compare to her speech across factors, she noted ``\textit{Ahhh... much more pose diversity than me.}'' With the Mirror Panel EA2 was quickly able to find speakers that use gestures very differently, more often over the border of the screen, and with a wider variety.

\section{Discussion}

In this section, we discuss the lessons learned in our research, the limitations of our work, and the implications for future work.

\textbf{Analysis of Online Public Speaking.} 
Our system has received high appreciation from domain users for its ability to evaluate and analyze online speech techniques. This demonstrates the value and potential of our system.
Users have also expressed their expectation for additional features and capabilities in our system. Most of the participants mentioned that a report of the analysis results can offer an intuitive understanding of a speech, alleviate the analysis burden for users, and provide 
clear guidance for future exploration. 
Our system can be further simplified for better user experience by putting less on the interface.
Improving text visualization and analysis is a potential area for future work, addressing \textit{SpeechMirror}'s limitations in readability and supporting advanced analysis. This can be achieved through exploring text feature analysis and incorporating text visualizations \cite{kucher2015text}.

\textbf{Visualization of Human Body Language.}
We have employed novel approaches to visualize data in different modalities of presentation techniques. Combining multiple modalities of data to provide a unified visualization analysis while preserving the distinctive characteristics of each modality poses significant challenges. Balancing the amount of information and complexity in multimodal visualization requires further consideration. 
Additionally, the color-emotion mapping scheme in our system may hinder visual accessibility, particularly for individuals with red-green color blindness. To address this, an optional color scheme is provided in Section 4.3 of the supplementary material, enabling individuals with color blindness to utilize our system.

\textbf{Scope of Presentation Techniques.} We are compiling a comprehensive list of speech techniques for users to explore and improve, 
and it's been positively received.
However, domain experts have suggested more techniques to consider. 
Our evaluation revealed that background scenery, lighting, and visual aids are crucial for online speeches. 
Extracting these features requires further investigation.
Initial attempts to detect prop usage via human-object interaction models \cite{zhang2021scg,zhang2022upt} had suboptimal outcomes due to the limited recognizable objects and interaction classes. 
As relevant models advances, we expect more presentation techniques to be quantified and analyzed in future work.
Expanding on body gestures in our work, advances in hand pose estimation \cite{deng2023recurrent,tu2023consistent} enable the analysis of hand gestures in public speaking.

\textbf{Recommendation of Speeches.}
The recommendation function in our system simplifies the process of discovering new videos for users, and has been well-received by our participants. More recommendation approaches can be further considered, such as based on the themes or topics of speeches, on the placements of speeches, or on other tags for users to filter. Moreover, recommendations made by fusing multimodal raw features of presentation techniques could be promising.

\textbf{Model Accuracy.} 
We utilized state-of-the-art models for multimodal feature extraction. However, apart from the aforementioned limitations regarding the extraction of presentation techniques, we observed inaccuracies of the models in certain situations. For instance, when a webcam is positioned too low, capturing an upward-facing view of the speaker's face, it can result in inaccurate eye gaze and facial emotion recognition. Overall, our models meet practical accuracy requirements, but the performance can be enhanced by incorporating more precise feature extraction models. For the factor effectiveness estimation model, we may explore more precise models or models that consider multiple factors simultaneously. 
To address potential inaccuracies, we suggest introducing uncertainty visualization in future research. Accurately assessing and visually encoding uncertainty may pose significant challenges.

\textbf{Volume of Data.} \textit{SpeechMirror} is the first visual analytics system that analyzes the effectiveness of speeches and recommends speech samples based on a collection of videos. 
We anticipate that as the volume of speech video data increases, our effectiveness estimation model will generate more accurate and robust results.
With larger volume of video, the system will also provide more diverse recommendation results in both expression and speech content.

\textbf{Generalization of Our Methods.} Our work focuses on online 
speech contests in the WCPS. Our work can also be extended to other speaking and presentation scenarios in the future.
Furthermore, one participant in the evaluation (EE1) suggested that our system could be adapted and applied to other fields that require the analysis of body language, such as behavior analysis during suspect interrogations.

\textbf{Privacy Issue.} \textit{SpeechMirror} provides users with other speakers' speech videos as a basis for analysis, which inevitably raises privacy concerns for the speakers. Therefore, we obtain the consent of the speakers in actual usage and restrict users' access to the data and system. However, when deploying similar systems in public use, ethical considerations should be addressed. It is also important to consider the potential for imitation and plagiarism of presentation techniques with such systems, which requires further exploration in future work.

\section{Conclusion}

In this paper, we propose \textit{SpeechMirror}, a visual analytics system for public speaking experts and amateurs to evaluate a speech and explore potential improvements.
Our system is the first to evaluate speaker performance on a single speech
for speaking techniques and provide an in-depth analysis using a comprehensive set of factors, allowing users to identify and understand areas for improvement of their speech. 
Additionally, the system recommends users speeches by similarity and difference of techniques and content to enhance their understanding of different ways of expression. 
Novel visualizations are employed for intuitive understanding of presentation techniques.
We designed the system based on insights from literature review and domain interviews. 
A user study was conducted to evaluate the system, indicating the effectiveness of our system in user experience and gaining insights.

In future work, we plan to further improve our system in usability and functionality based on the aforementioned reflections on our system, including (1) making the interface easier to use, (2) expanding the scope of presentation techniques, (3) improving the accuracy of models. Furthermore, we intend to promote our system for applications and collect more speech data for more potential research.

\acknowledgments{
We wish to thank our domain collaborators and the anonymous reviewers for their constructive comments.
This work was supported by the National Key R\&D Program of China (2022ZD0117900) and Beijing Natural Science Foundation (4212029).
}

\bibliographystyle{abbrv-doi-hyperref}

\bibliography{template}
\end{document}

\maketitle

\section{Details of the Design Process}

In this section, we will introduce the domain-centered interviews of \textit{SpeechMirror} in detail. 

\subsection{Details of participants in domain interview}

The details of participants in domain interview are listed in \autoref{tab:participant-domaininterview}.

\textbf{Basic Information.} All the experts are male while both amateurs are female. 
The age ranges for both experts and amateurs are similar, with both groups having participants in 18-50 years old. There is also an expert in the 51-60 age range. 
Two expert participants speak English as their native language, while other participants speak Chinese.

\textbf{Educational Background.} All amateur participants are with non-STEM backgrounds, while expert participants have both STEM and non-STEM backgrounds.
Regarding the participants' educational background, only two individuals possess relevant academic training. Specifically, one participant has a background in mathematics (statistics) and computer science, while the other has a background in psychology. The remaining participants have educational backgrounds that are not directly relevant to our study.

\textbf{Public Speaking Experience. }All experts had competed in over 5 contests, while both amateurs have competed in 2-5 times.
For online contests, one expert have competed for only once, one expert for 2-5 times and two experts for more than 5 times. The two amateurs have participated in contests once and 2-5 times respectively.
Among all participants, one amateur participant has watched 1-10 speeches, while one expert and one amateur participant have watched 11-50 speeches. The remaining participants have watched more than 50 speeches.
Experts have participants who have coached for 0-1 (1 participant) and 3-5 years (3 participants), while amateurs have no experience of training others as an occupation.

\begin{table*}[tb]
\caption{The information of participants in domain interview}
\label{tab:participant-domaininterview}
\resizebox{\linewidth}{!}{
\begin{tabular}{@{}llllllllll@{}}
\toprule
Participant ID & Gender & Age        & STEM/Non-STEM & Educational background                                                             & \begin{tabular}[c]{@{}l@{}}Native\\ language\end{tabular} & \begin{tabular}[c]{@{}l@{}}Contests\\ competed\end{tabular} & \begin{tabular}[c]{@{}l@{}}Online contests \\ competed\end{tabular} & \begin{tabular}[c]{@{}l@{}}Speeches \\ watched\end{tabular} & \begin{tabular}[c]{@{}l@{}}Years of training \\ as occupation\end{tabular} \\ \midrule
DE1             & Male   & 31$-$40 & Non-STEM      & None                                                                               & English                                                   & over 5                                                      & over 5                                                              & over 50                                                     & 3$-$5                                                                   \\

DE2             & Male   & 51$-$60 & STEM          & \begin{tabular}[c]{@{}l@{}}Mathematics (Statistics)\\ Computer Science\end{tabular} & English                                                   & over 5                                                      & 1                                                                   & over 50                                                     & 3$-$5                                                                   \\
DE3             & Male   & 41$-$50 & Non-STEM      & Psychology                                                                         & Chinese                                                   & over 5                                                      & 2-5                                                                 & over 50                                                     & 3$-$5                                                                   \\
DE4             & Male   & 31$-$40 & STEM          & None                                                                               & Chinese                                                   & over 5                                                      & over 5                                                              & 11-50                                                       & 0$-$2                                                                   \\
DA1             & Female & 18$-$25 & Non-STEM      & None                                                                               & Chinese                                                   & 2-5                                                         & 2-5                                                                 & 11-50                                                       & No                                                                         \\
DA2             & Female & 26$-$30 & Non-STEM      & None                                                                               & Chinese                                                   & 2-5                                                         & 1                                                                   & 1-10                                                        & No                                                                         \\ \bottomrule
 
\end{tabular}
}
\end{table*}

\subsection{Details of the process of domain interview}

The interviews mainly focused on four topics in the form of questions:

\textit{Q1: What presentation techniques might be especially important for online public speaking? How they affect the performance of public speaking?}

\textit{Q2: What presentation techniques are more critical than others in online public speaking? Please list the top 8 techniques and explain why.}

\textit{Q3: How do you usually analyze and improve your speeches? How do you learn from others' speeches?}

\textit{Q4: Is there something that is not easy to observe or easy to ignore when watching speech videos? }

Participant answers to Q1 and Q2 helped us to better understand the scope of factors covered by our system, as well as the prioritization of the factors in the system. 

Q1 was a verbal question, meant to find if there were factors not considered by us in designing our system, and how the factor affects speech performance. 

For Q2 we provided 20 factors which the interviewer would rank according to their view of the factors' importance in speeches. Among the top 4 factors were emotional diversity, eye gaze, emotional intensity, and emotional coherence.

 Q3 and Q4 helped to establish the analytical tasks the system is meant to accomplish.

Live feedback for online speeches is particularly challenging, as DA1 stated that ``\textit{faces of the audience are not shown, and no audio reply is allowed. You don’t know if the audience gets your points. . . and can’t change without feedback.}''

\section{Details of the Data Collection}
\begin{figure}[tb]
 \centering 
 \includegraphics[width=\linewidth]{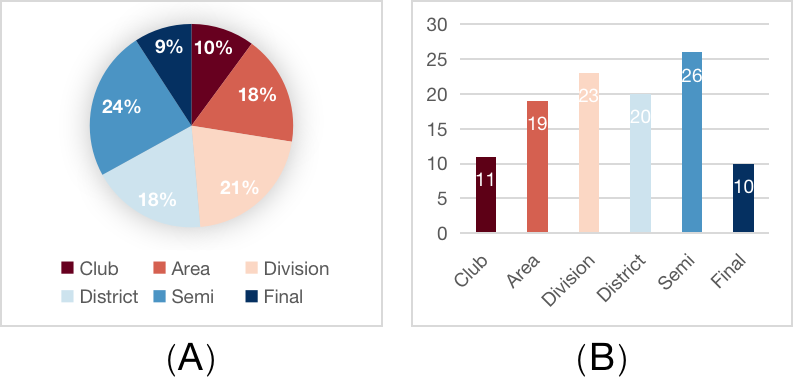}
 \caption{The detail information of the data collection}
 \label{fig:sp_dataCollection}
\end{figure}

While collecting speech videos of over 100 online speech videos from open access channels, we tried to keep a balance between quality and quantity. More details of data statistical distribution are shown in \autoref{fig:sp_dataCollection}.

\section{Details of Factor Analysis Result}
In ordinal regression, we perform logistic regression on the odds ratio
of each factor. The result of ordinal regression is shown in \autoref{tab:ordinal-regression}.
\begin{table*}[tb]
\centering
\caption{The result of ordinal regression}
\label{tab:ordinal-regression}
\resizebox{\linewidth}{!}{
\begin{tblr}{
  cell{2}{1} = {r=5}{},
  cell{3}{2} = {r=2}{},
  cell{5}{2} = {r=2}{},
  cell{7}{1} = {r=3}{},
  cell{7}{2} = {r=2}{},
  cell{10}{1} = {r=4}{},
  cell{10}{2} = {r=2}{},
  cell{12}{2} = {r=2}{},
  cell{14}{1} = {r=3}{},
  cell{14}{2} = {r=2}{},
  cell{17}{1} = {r=4}{},
  cell{17}{2} = {r=2}{},
  cell{19}{2} = {r=2}{},
  cell{21}{1} = {r=4}{},
  cell{21}{2} = {r=2}{},
  cell{23}{2} = {r=2}{},
  cell{25}{3} = {c},
  cell{25}{4} = {c},
  cell{25}{5} = {c},
  cell{25}{6} = {c},
  cell{25}{7} = {c},
  cell{25}{8} = {c},
  cell{25}{9} = {c},
  cell{25}{10} = {c},
  hline{1-2,7,10,14,17,21,25-26} = {-}{},
  hline{3,5,9,12,16,19,23} = {3-10}{},
}
Presentation Technique & Feature              & Factor     & Significance & w        & $b_0$  & $b_1$  & $b_2$   & $b_3$  & $b_4$  \\
Facial Expression      & Type                 & Diversity  & $p=0.002^*$  & -2.028   & 0.318  & 1.262  & 2.361   & 3.207  & 4.87   \\
                       & Valence              & Volatility & $p=0.571$    & -0.017   & -2.626 & -1.751 & -0.709  & 0.103  & 1.721  \\
                       &                      & Average    & $p=0.005^*$  & 3.698    & -2.249 & -1.322 & -0.238  & 0.581  & 2.257  \\
                       & Arousal              & Volatility & $p=0.761$    & -0.009   & -2.41  & -1.534 & -0.496  & 0.31   & 1.926  \\
                       &                      & Average    & $p=0.000^*$  & 12.71    & -1.311 & -0.343 & 0.846   & 1.786  & 3.559  \\
Eye Contact            & Gaze Direction       & Volatility & $p=0.002^*$  & -597.981 & -4.139 & -3.162 & -2.012  & -1.187 & 0.431  \\
                       &                      & Dispersion & $p=0.067$    & 5.245    & -1.082 & -0.189 & 0.87    & 1.686  & 3.313  \\
                       & Watching Camera      & Ratio      & $p=0.265$    & 1.536    & -1.923 & -1.03  & 0.014   & 0.814  & 2.429  \\
Use of Stage           & Distance from Camera & Volatility & $p=0.908$    & 0.001    & -2.095 & -1.216 & -0.18   & 0.624  & 2.238  \\
                       &                      & Dispersion & $p=0.185$    & 1.444    & -1.903 & -1.021 & 0.023   & 0.837  & 2.476  \\
                       & Position in Frame    & Volatility & $p=0.026^*$  & -210.54  & -2.957 & -2.044 & -0.963  & -0.144 & 1.493  \\
                       &                      & Dispersion & $p=0.141$    & 0.006    & -1.59  & -0.71  & 0.332   & 1.146  & 2.787  \\
Body Gesture           & Gesture Energy       & Volatility & $p=0.860$    & 0.005    & -1.962 & -1.082 & -0.045  & 0.757  & 2.371  \\
                       &                      & Average    & $p=0.426$    & 4.28E-07 & -2.041 & -1.164 & -0.128  & 0.68   & 2.306  \\
                       & Gesture Diversity    & Diversity  & $p=0.266$    & 191.157  & -1.833 & -0.945 & 0.099   & 0.907  & 2.527  \\
Voice                  & Volume               & Volatility & $p=0.000^*$  & -0.17    & -8.904 & -7.547 & -6.055  & -5.13  & -3.444 \\
                       &                      & Average    & $p=0.413$    & -0.048   & -4.954 & -4.071 & -3.035  & -2.233 & -0.612 \\
                       & Pitch                & Volatility & $p=0.438$    & -0.012   & -3.088 & -2.207 & -1.165  & -0.359 & 1.257  \\
                       &                      & Average    & $p=0.988$    & 0.000071 & -2.103 & -1.224 & -0.187  & 0.616  & 2.229  \\
Pace                   & Speaking Rate        & Volatility & $p=0.617$    & 0.023    & -1.254 & -0.37  & 0.668   & 1.47   & 3.084~ \\
                       &                      & Average    & $p=0.198$    & 0.007    & -0.286 & 0.607  & 1.668   & 2.473  & 4.076  \\
                       & Pauses               & Volatility & $p=0.157$    & 0.065    & 0.492  & 1.39   & 2.44    & 3.248  & 4.872  \\
                       &                      & Average    & $p=0.533$    & -0.002   & -2.416 & -1.53  & -0.489~ & -0.314 & 1.927  \\
Content                & Script               & -          & -            & -        & -      & -      & -       & -      & -      
\end{tblr}}
\end{table*}

\section{Details of the visual design}

\subsection{Design alternatives}

During the creation of \textit{SpeechMirror} there were many design alternatives that were created. We iterated our designs using earlier versions of system to test potential designs.

\subsubsection{Consistently used icons}

\begin{figure*}[tb]
 \centering 
 \includegraphics[width=\linewidth]{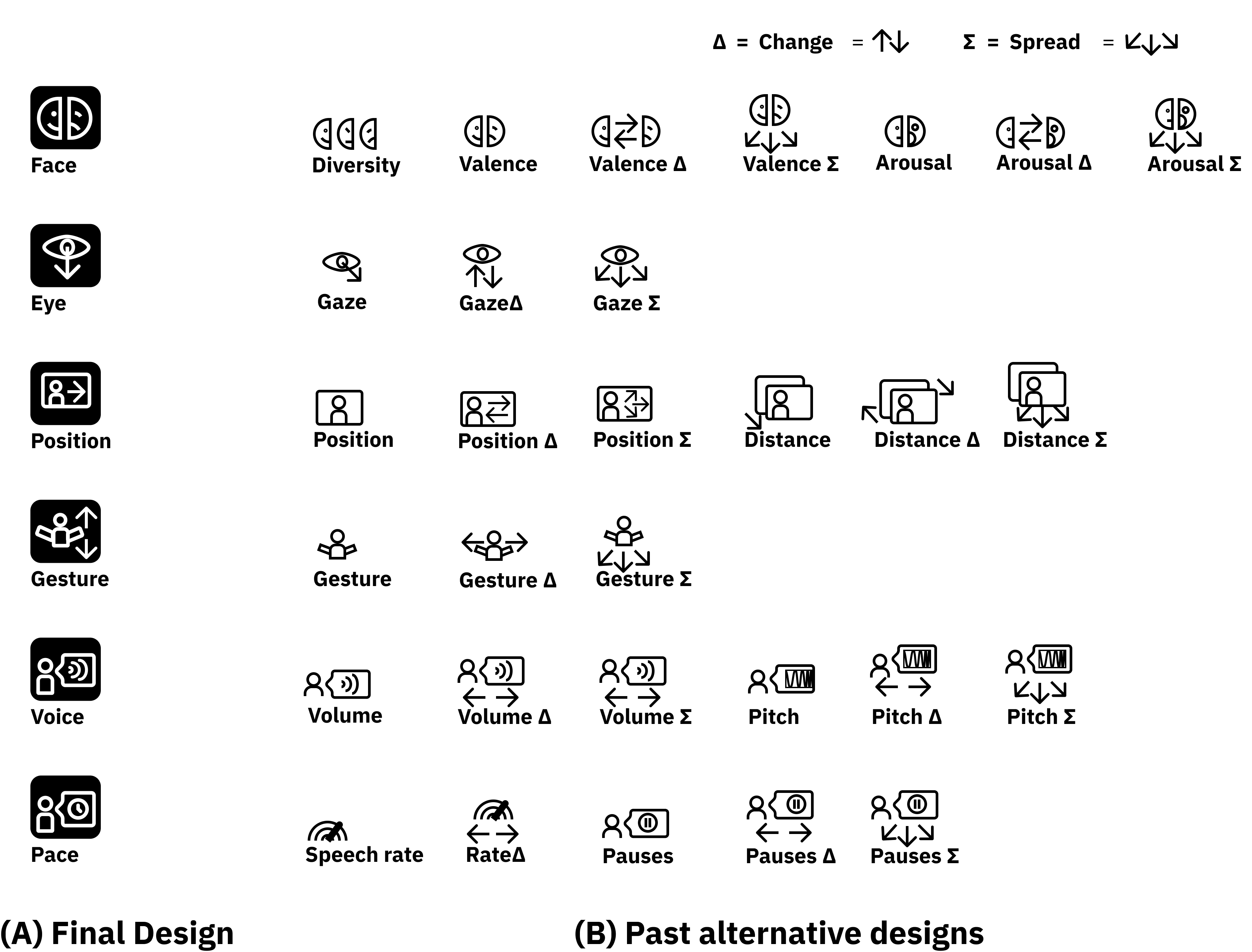}
 \caption{Design alternatives of the consistently used icons.}
 \label{fig:icon-alternative}
\end{figure*}

Early on in the design we decided to provide users with more consistently used icons in the system that would provide a more explanatory overview of their speech. However, there remains uncertainty in the complexity of the icons.

Our first attempt was to provide as comprehensive as possible view to the user of the many factors considered by our system, as seen in the past alternative design \autoref{fig:icon-alternative} (B). 
A broader scope of factors was considered, conceptually showing the change of factors (volatility) and spread (dispersion) of factors through arrows that suggest their purpose.
In theory this would allow users to have a common visual language to more easily understand these concepts. However, the first time we evaluated the system with this set of icons, users had much confusion. The visual complexity of the icons, as well as the broader scope of the factors used in our system seemed to create unnecessary confusion. After this first unsatisfactory evaluation, we then decided to simplify the visual language used in the system, using a smaller and more memorable set of icons, as seen in \autoref{fig:icon-alternative} (A). 

\subsubsection{SpeechTwin and the Mirror Panel}

\begin{figure*}[tb]
 \centering 
 \includegraphics[width=\linewidth]{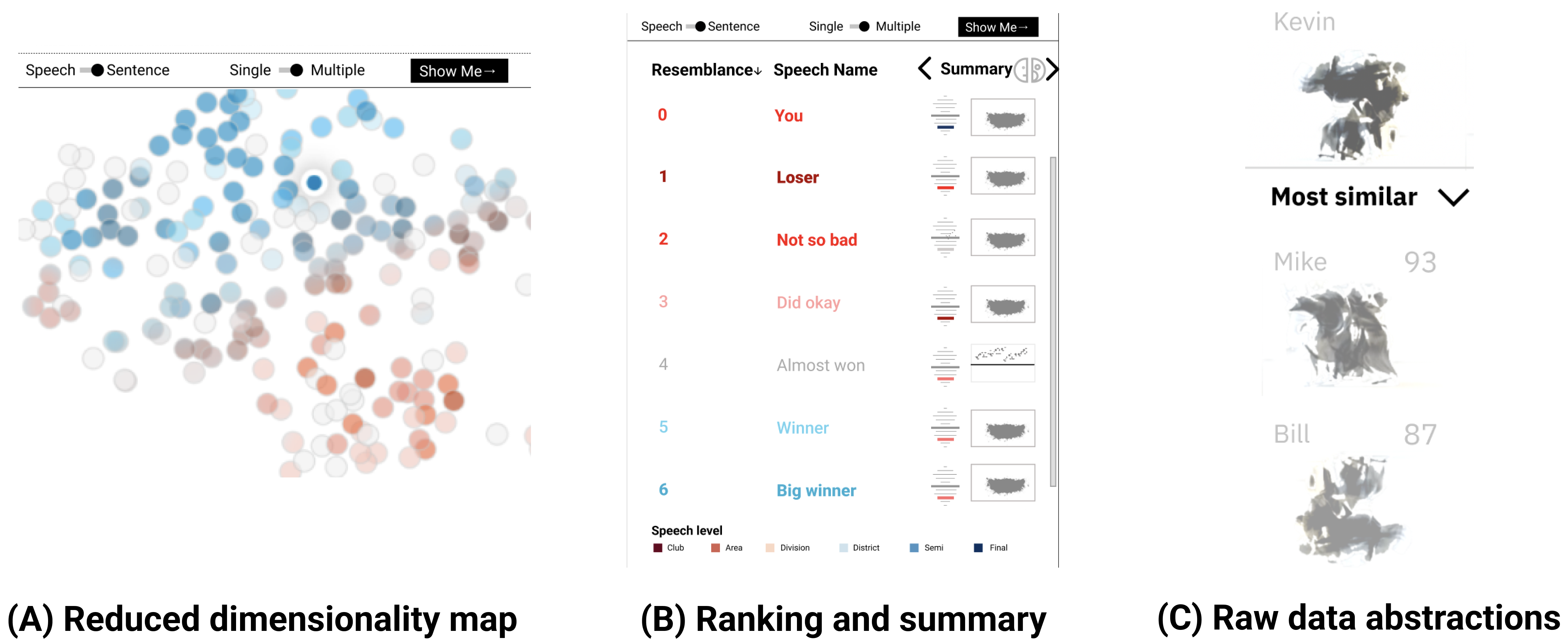}
 \caption{Design alternatives of SpeechTwin and the Mirror Panel.}
 \label{fig:mirror-panel-alternative}
\end{figure*}

With the purpose of giving users a visual summary of the similarity between their speech and other speeches as well as an understanding of how crucial speaking techniques in their speech relate to other speeches, we attempted several different visualizations. In the past work E-ffective \cite{effective2022}, the authors created a visualization called E-similarity that sought to achieve a similar purpose ``to compare the similarity between speeches'' as well as provide a visual summary of the speech’s most crucial factors. E-similarity compared more similar speeches as more close in distance on a two-dimensional map. Each speech was represented as a point on the map, which used t-SNE \cite{van2008visualizing} to reduce dimensionality. Upon clicking a speech, a radar chart showing the most crucial speaking techniques was displayed. One design alternative we created was similar to E-similarity, but used color encoding and icons used elsewhere in our system, which can be seen in \autoref{fig:mirror-panel-alternative} (A). Informal sessions to test the design of this visualization suggested that it was difficult for the audience of our system to understand how the map encodes similarity. This is similar to the result of the E-ffective evaluation, where the users rated the ease of user of e-similarity poorly. 

An alternative design provided a ranking of speeches according to similarity, as shown in \autoref{fig:mirror-panel-alternative} (B). The advantage of this design is that it provided a more readily understandable list. The disadvantage included that it poorly used the space, and the summary of crucial factors was only observable by navigation. In our final design, we kept the method of ranking by similarity in the mirror panel, but provided information of crucial factors in the glyph provided, SpeechTwin. 

Another design direction we considered was showing views of the raw data over time in small multiple views, see \autoref{fig:mirror-panel-alternative} (C). Inspiration came from the work of Diezmann et al. \cite{diezmann2003grids}.
An implementation of their methods can be seen in in the work of Gremmler \cite{gremmler2016kung}
that similarly shows the volatility of motion across time. However, we found the resulting visualizations difficult to achieve for all the factors our system considered. Additionally, displaying each factor independently also had the disadvantage of not allowing users to get a summary of the most critical factors to compare.

\begin{figure*}[tb]
 \centering 
 \includegraphics[width=\linewidth]{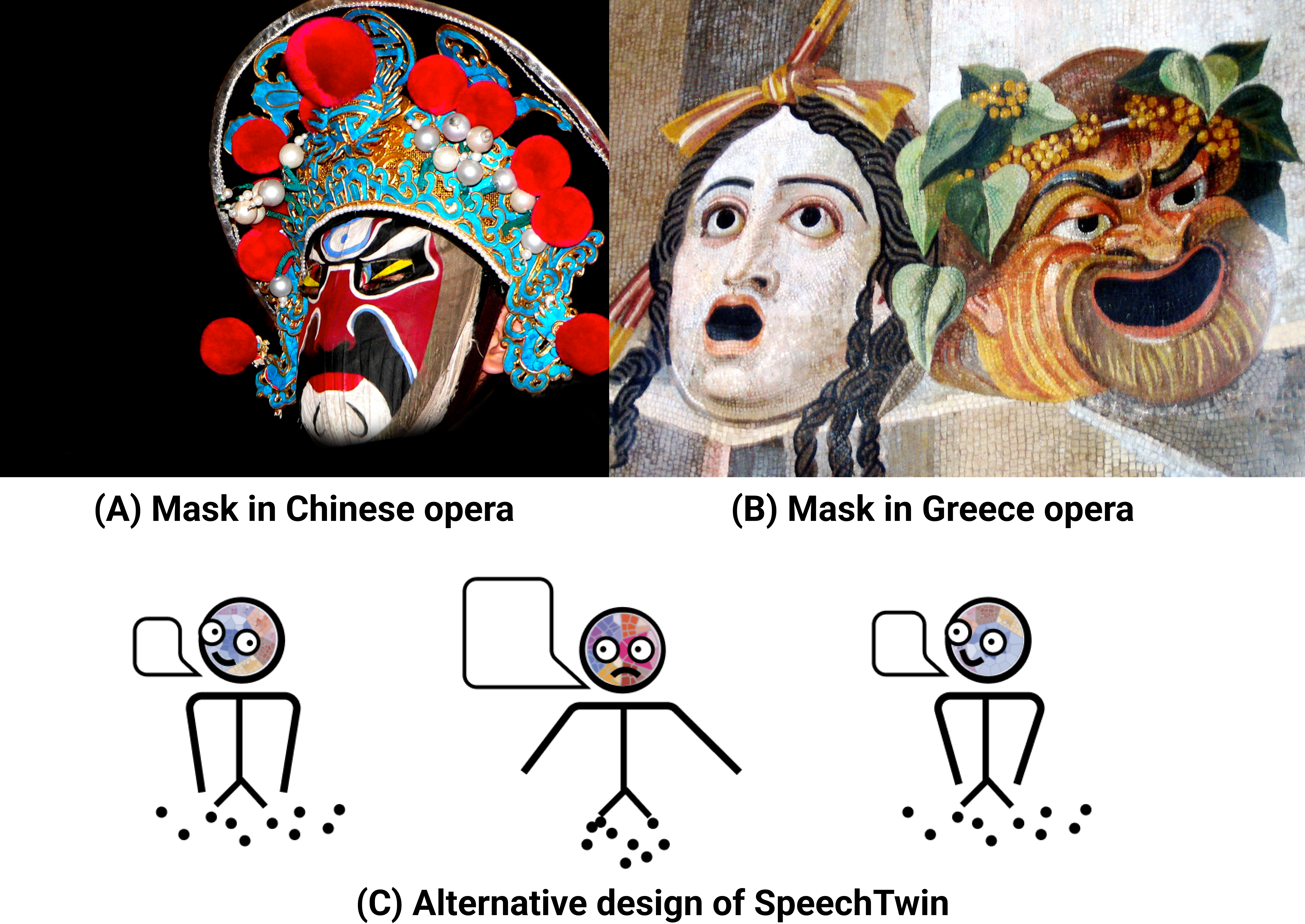}
 \caption{Design inspirations and alternatives of SpeechTwin. (A) A mask in Chinese opera, refer \cite{yongxinge2007bianlian}. (B) A mask in Greece opera, refer \cite{unknown200tragic}. (C) A design alternative of SpeechTwin.}
 \label{fig:speechtwin-alternative}
\end{figure*}

For the creation of the visualization used in our final system, SpeechTwin, we were inspired by the masks used in different cultures to give a summary of the qualities of the speaker. In Chinese operas, masks that reflect the mood and personality of the roles they are performing are commonly used, see \autoref{fig:speechtwin-alternative} (A) for an example. In Greek theater, the masks provide exaggerated summaries of the roles of the speaker through changes such as the shape of the mouth and shape of the facial muscles, see \autoref{fig:speechtwin-alternative} (B). 

As illustrated in \autoref{fig:speechtwin-alternative} (C), alternative designs for SpeechTwin included showing the proportion of emotions as Veronoi diagrams contained in the boundaries of face, as well as speech bubbles that illustrate qualities of the speaker’s voice.

\subsection{Demonstration of data visualization}

\begin{figure*}[tb]
 \centering 
 \includegraphics[width=\linewidth]{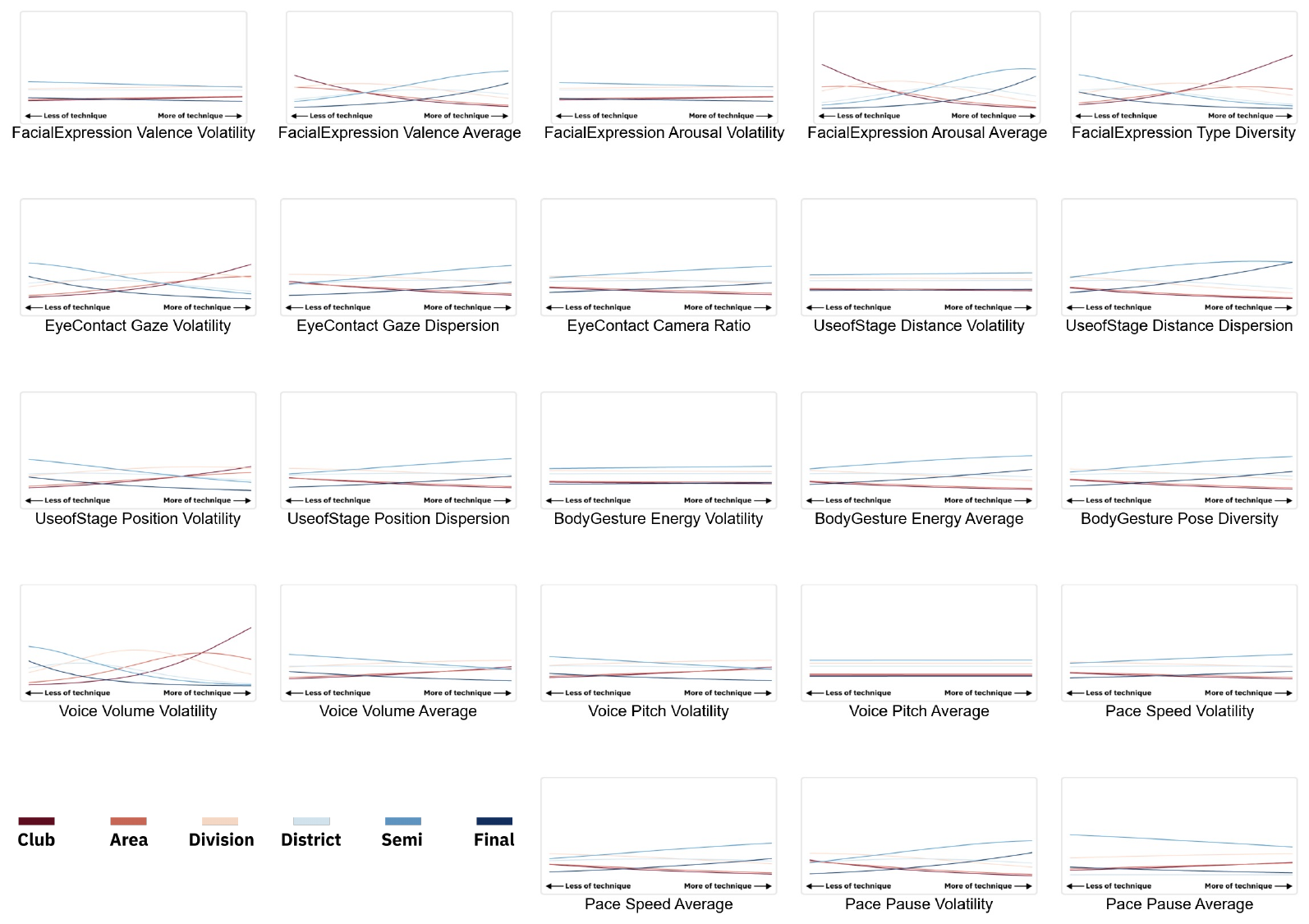}
 \caption{Effective curve.}
 \label{fig:effective-demo}
\end{figure*}

\begin{figure*}[tb]
 \centering 
 \includegraphics[width=\linewidth]{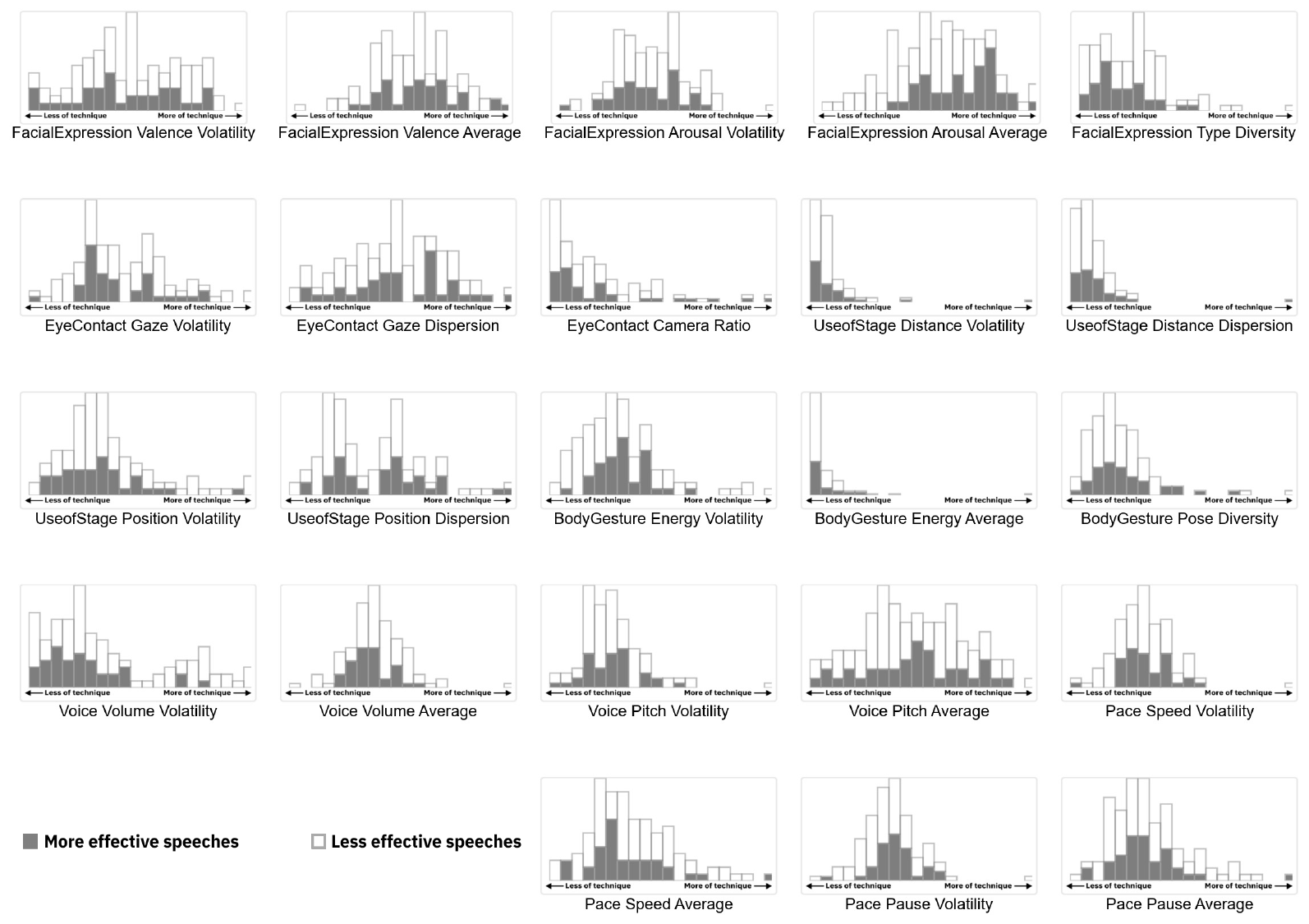}
 \caption{Factor Histogram.}
 \label{fig:hist-demo}
\end{figure*}

\begin{figure*}[tb]
 \centering 
 \includegraphics[width=\linewidth]{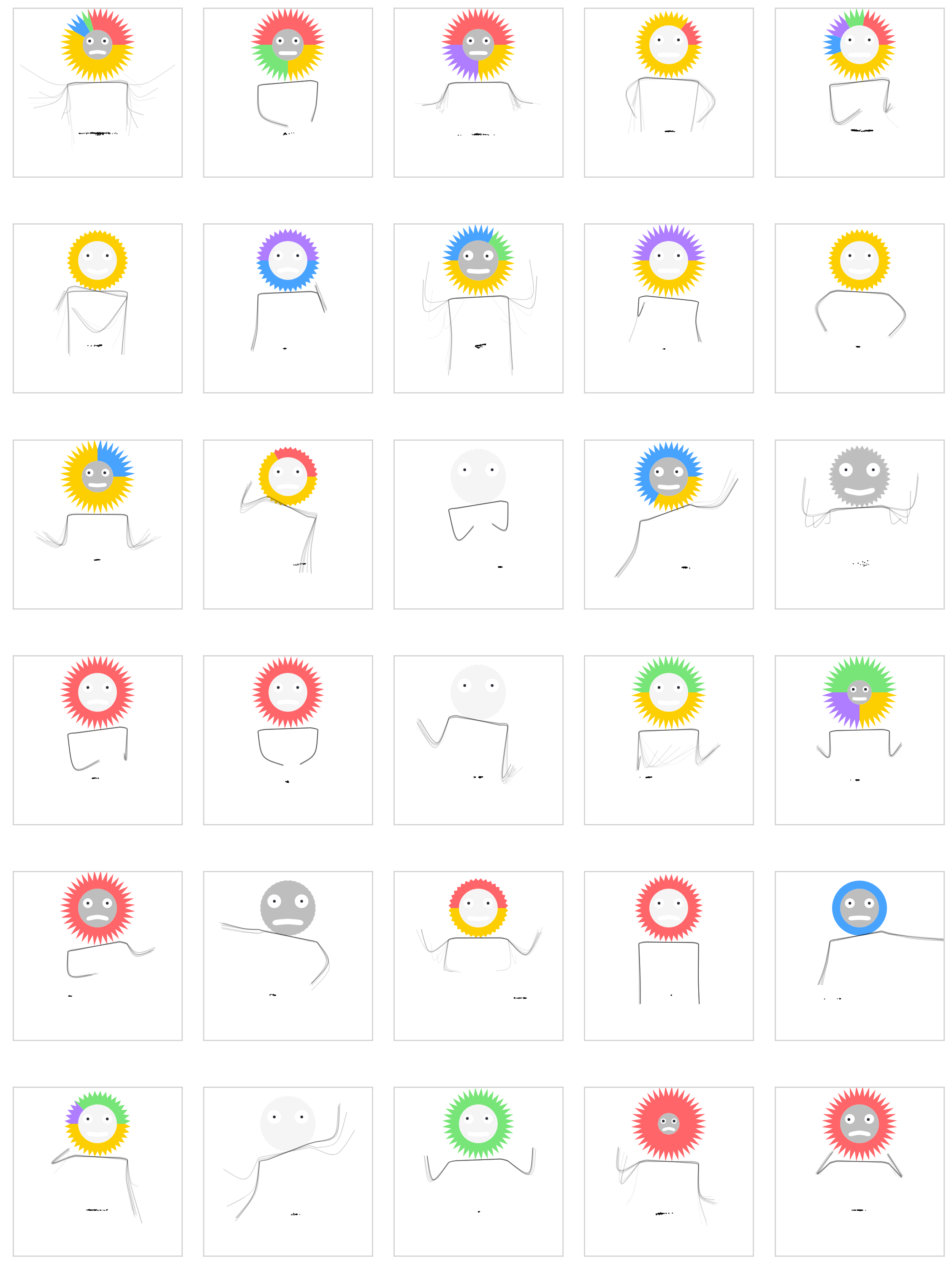}
 \caption{SpeechTwins of sentences.}
 \label{fig:SpeechTwins-sentences}
\end{figure*}

\begin{figure*}[tb]
 \centering 
 \includegraphics[width=\linewidth]{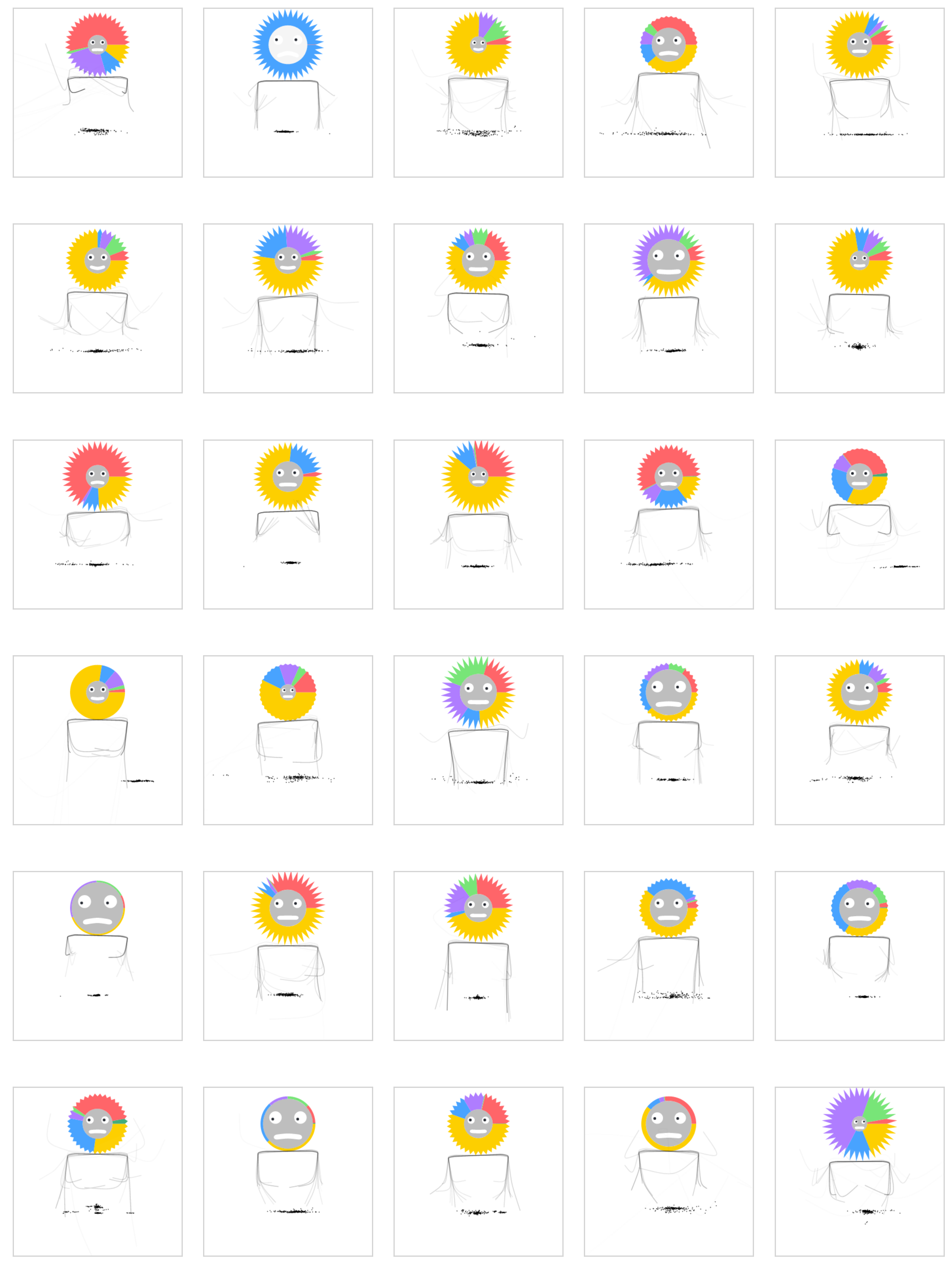}
 \caption{SpeechTwins of speeches.}
 \label{fig:SpeechTwins-speeches}
\end{figure*}

In this section, we provide more detailed visualization demos for our proposed visualization methods: Effective curve (\autoref{fig:effective-demo}), Factor Histogram (\autoref{fig:hist-demo}), SpeechTwins of sentences (\autoref{fig:SpeechTwins-sentences}) and SpeechTwins of speeches (\autoref{fig:SpeechTwins-speeches}).

\subsection{Accessibility of data visualization}

\begin{figure*}[htbp]
 \centering 
 \includegraphics[width=\linewidth]{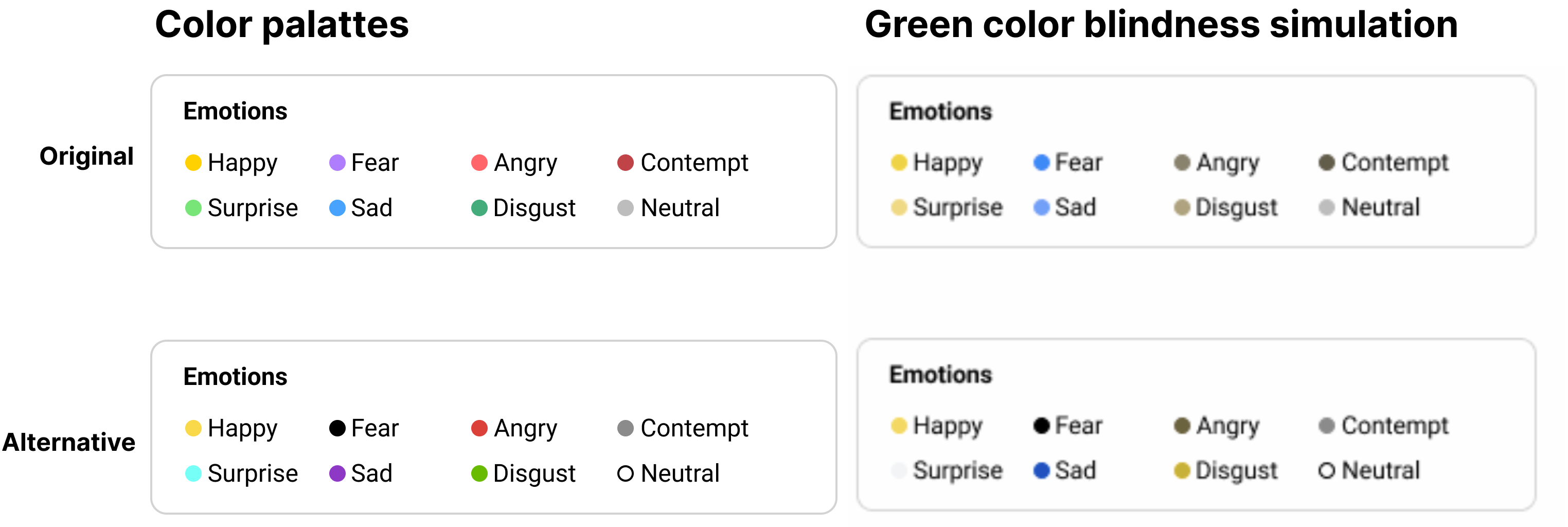}
 \caption{Alternative color encoding palette and color blindness simulation result.}
 \label{fig:color-alternative}
\end{figure*}

\begin{figure*}[htbp]
 \centering 
 \includegraphics[width=\linewidth]{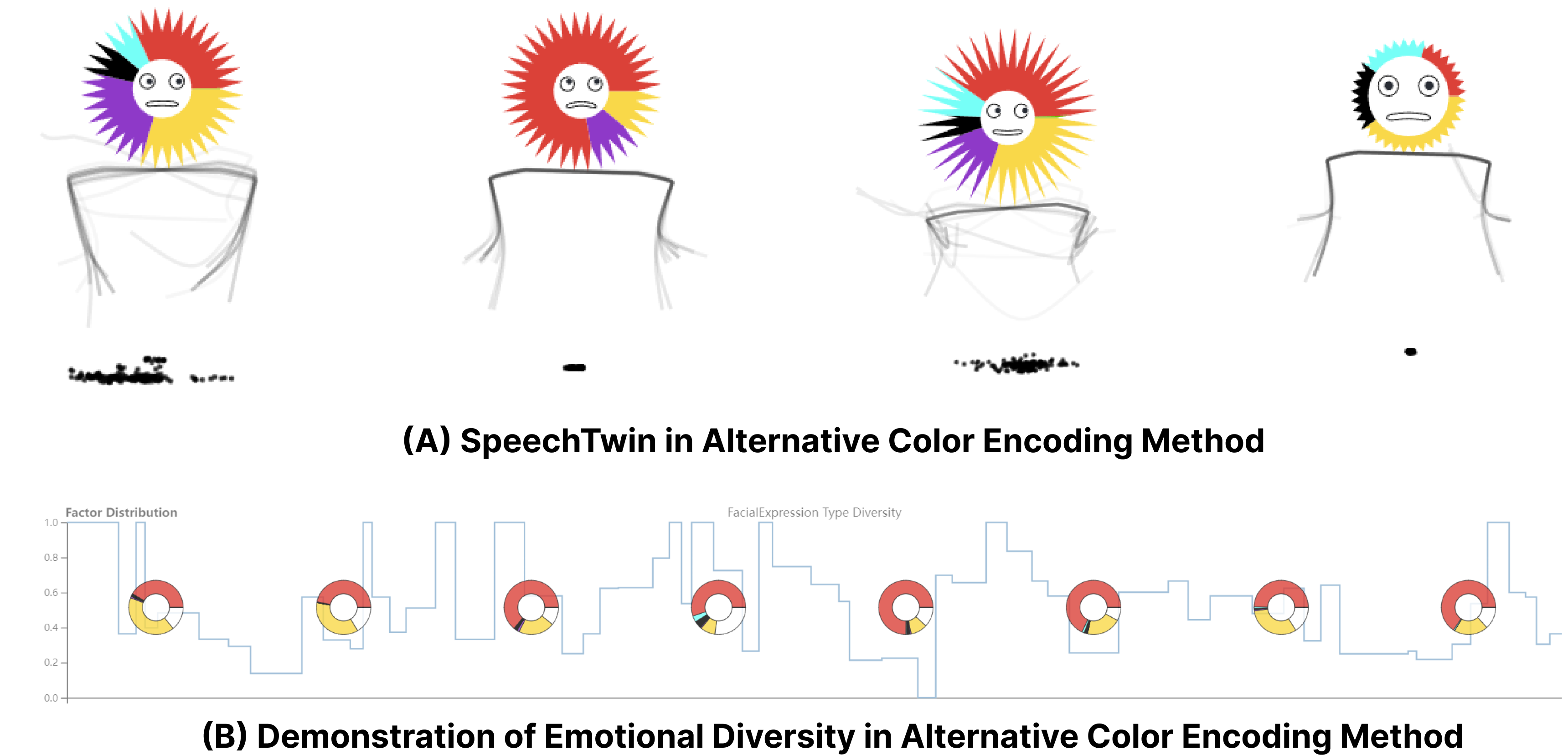}
 \caption{Demonstrations of visualizations with the alternative color encoding palette. (A) SpeechTwin. (B) Emotional Diversity visualization.}
 \label{fig:color-alternative-viz}
\end{figure*}

For red and green color blind people, the original colors we used to identify emotional types proved problematic by color blindness simulation. In order to provide access for individuals with color blindness, an alternative color encoding method was created. Jonauskaite et al. \cite{jonauskaite2021colour} evaluated the emotional associations of different colors for color blind and non-color blind individuals. We referenced their work when deciding on the colors associated with our revised color palette. We then used an online color blindness simulator \cite{colorblindsimulator} developed with methods by 
Machado et al. \cite{machado2009a} in order to further differentiate the colors for the more common red-green colorblind individuals. The resulting color set, as well as the simulated colors can be seen in \autoref{fig:color-alternative}. The visualizations in the alternative color encoding method are shown in \autoref{fig:color-alternative-viz}.

\section{Details of the Evaluation Study}

\subsection{Details of participants}

The details of the participants in our evaluation study are listed in \autoref{tab:participant-evaluation}.

\begin{table*}[tb]
\caption{The information of participants in evaluation}
\label{tab:participant-evaluation}
\resizebox{\linewidth}{!}{
\begin{tabular}{@{}llllllllll@{}}
\toprule
Participant ID & Gender & Age   & STEM/Non-STEM & Educational backgrond                                                                           & \begin{tabular}[c]{@{}l@{}}Native \\ language\end{tabular} & \begin{tabular}[c]{@{}l@{}}Contests \\ competed\end{tabular} & \begin{tabular}[c]{@{}l@{}}Online contests \\ competed\end{tabular} & \begin{tabular}[c]{@{}l@{}}Speeches\\ watched\end{tabular} & \begin{tabular}[c]{@{}l@{}}Years of training\\ as occupation\end{tabular} \\ \midrule
EE1            & Male   & 26-30 & STEM          & None                                                                                            & Chinese                                                    & over 5                                                       & 2-5                                                                 & over 50                                                    & over 5                                                                    \\
EE2            & Female & 31-40 & STEM          & \begin{tabular}[c]{@{}l@{}}Psychology\\ Mathematics (Statistics)\\ Computer Science\end{tabular} & Chinese                                                    & over 5                                                       & 2-5                                                                 & over 50                                                    & 3-5                                                                       \\
EE3            & Female & 31-40 & Non-STEM      & None                                                                                            & Chinese                                                    & over 5                                                       & 2-5                                                                 & over 50                                                    & over 5                                                                    \\
EE4            & Male   & 41-50 & STEM          & \begin{tabular}[c]{@{}l@{}}Mathematics (Statistics)\\ Computer Science\end{tabular}              & Other                                                      & over 5                                                       & over 5                                                              & over 50                                                    & 3-5           \\
EA1            & Female & 31-40 & Non-STEM      & None                                                                                            & Chinese                                                    & 2-5                                                          & 2-5                                                                 & 11-50                                                      & No                                                                        \\
EA2            & Female & 31-40 & Non-STEM      & None                                                                                            & Chinese                                                    & over 5                                                       & 2-5                                                                 & over 50                                                    & No                                                                       \\
EA3            & Male   & 51-60 & Non-STEM      & Design                                                                                          & Chinese                                                    & over 5                                                       & over 5                                                              & 11-50                                                      & No                                                                        \\
EA4            & Female   & 31-40 & STEM      & Computer Science                                                                                          & Chinese                                                    & 2-5                                                       & 2-5                                                              & 11-50                                                      & No                                                                        \\

\\ \bottomrule
\end{tabular}
}
\end{table*}

\subsection{Appratus of the study}

The study was conducted remotely, utilizing video conferencing software to control the system interface. The participants were required to keep their webcam on and the interface set to full screen throughout the evaluation process.

\subsection{Details of user task}
The user tasks of participants in evaluation are listed in \autoref{tab:user-task}.

\begin{table*}[tb]
\centering
\caption{The details of user task in evaluation}
\label{tab:user-task}
\resizebox{\linewidth}{!}{
\begin{tblr}{
  hline{1,14} = {-}{0.08em},
  hline{2} = {-}{0.05em},
}
ID   & User Task        & Corresponding Design Considerations                                  \\
UT1  & Find a factor our system thinks might improved in the speech.    &         DC1              \\
UT2  & Evaluate how the speech factor is used in the world final and semi-final speeches.  & DC1   \\
UT3  & Try to understand the effectiveness distribution of the factor.    &  DC1                \\
UT4  & Click to select the factor to focus on the use of the technique in the speech.  &  DC3, DC4 \\
UT5  & Play the speech and watch the delivery of the factor.         &  DC4                   \\
UT6  & In the video feed try to hover on the part of speaker in you are interested in. &  DC4     \\
UT7  & Find a potentially ineffective moment in the speech and select it.       &   DC3, DC4       \\
UT8  & Observe how the factor is used in the content of the speech.     &        DC4              \\
UT9  & Observe the speech twin for the original speech.               &    DC5                 \\
UT10 & Look for a speech that has very similar or different use of this factor than you.  &  DC2, DC5   \\
UT11 & Find a moment in the compared speech where the factor is used similarly or differently. &    DC4, DC5   \\
UT12 & Try to find a place in the original speech which might be improved.          &      DC1-DC5  
\end{tblr}}
\end{table*}

\subsection{Details of user experience questionnaire}
The questions and answers of participants in evaluation are listed in \autoref{tab:participant-design}.

\begin{figure*}[tb]
 \centering 
 \includegraphics[width=\linewidth]
 {figures/DetailsofCaseStudy.pdf}
 \caption{Procedure of Case Study.}
 \label{fig:Procedure-of-Case-Study}
\end{figure*}

\subsection{Details of case study}
The procedures of case study for expert user (EE1) and amateur user (EA2) are shown in \autoref{fig:Procedure-of-Case-Study}.

\subsubsection{Procedure of Case Study for Expert User (EE1)}
As shown in \autoref{fig:Procedure-of-Case-Study} (a), we introduce the detailed procedure of case study for expert user (EE1) as follows:

\textcircled{1} EE1 noticed from the Factor Panel that the speaker he was coaching had exceptionally low arousal.

\textcircled{2} EE1 then used the Time Slice Panel to view the detail of arousal.

\textcircled{3} EE1 then used the mirror panel to find one of the most different SpeechTwins. EE1 clicked on the SpeechTwin to set the speech as the comparison video. EE1 noted the speaker was the previous year’s world 2nd place winner.

\textcircled{4} EE1 used the timeline to find a moment of especially high arousal. After noticing the changes in arousal over time, EE1 said ``\textit{I think the tool could be very useful. In order to build a connection to the audience, you need contrast, you need the twists and turns of emotions. Then you can elicit emotions of the audience. I can use this to show my student that there is strong contrast [for this expert speaker]. I could let the student see where they are and compare with different speakers.}''

\textcircled{5} While EE1's mouse was hovering on a SpeechTwin with a smooth spread of positional footprints, EE1 added, ``\textit{With the SpeechTwin we can see where they are and compare with the other speakers to see possibilities.}''

\textcircled{6} EE1 then moved his mouse to indicate a SpeechTwin with a smooth distribution of footprints and commented ``\textit{this speaker had constant stage movement.}'' EE1 then pointed at a SpeechTwin with a clear 3 part distribution of footprints, stating ``\textit{if like this there is a clear stage design}''. EE1 noted the possible use of our system for coaching, ``\textit{Overall I feel like this is a tool that I could use and discuss with my students and find out different insights so that they can improve.}''

\subsubsection{Procedure of Case Study for Amateur User (EA2)}

As shown in \autoref{fig:Procedure-of-Case-Study} (b), we introduce the detailed procedure of case study for amateur user (EA2) as follows:

\textcircled{1} EA2 first noted that while overall, the average valence in her face was indicated by the system as performing well, the system found her speech most similar to someone that was smiling throughout their speech. EA2 guessed it might be that her sadness wasn't expressed obviously enough.

\textcircled{2} When looking at the Time Slice panel, EA2 observed that the ending of the speech was happy, ``\textit{This is correct.}'' However, EA2 then pointed to earlier time slices in her speech ``\textit{For these two parts I really wanted to show my sadness, my anger, but I guess I failed.}''

\textcircled{3} While comparing by selecting both the valence average and gesture energy change, EA2 stated ``\textit{I want to see who is the most similar to me.}''

\textcircled{4} EA2 selected the most similar speech, noting that the gestures, similar to hers, were mostly below the screen or on the screen border.

\textcircled{5} EA2 then looked at the Mirror Panel to find the most different speaker. Playing the video, EA2 saw a speaker that had many different gestures and stage changes. While looking at the speech comparison board to compare to her speech across factors, EA2 noted ``\textit{Ahhh... much more pose diversity than me.}''

\begin{table*}
\centering
\caption{The questions and answers of participants in evaluation}
\label{tab:participant-design}
\resizebox{\linewidth}{!}{
\begin{tblr}{
  cell{1}{9} = {t},
  hline{1,22} = {-}{0.08em},
  hline{2} = {1-8}{0.03em},
  hline{2} = {9}{},
}
Questions                                                                                                     & EE1 & EE2 & EE3 & EE4 & EA1 & EA2 & EA3 & EA4 \\
Did you find the system brought awareness to the overall use of factors in the analyzed speech?               & 6   & 5   & 6   & 7   & 5   & 5   & 7   & 7   \\
Were you able to observe new ways of speech delivery in speeches?                                             & 7   & 5   & 5   & 6   & 5   & 6   & 6   & 5   \\
Did you find the system recommended factor for improvement useful?                                            & 7   & 4   & 6   & 6   & 5   & 5   & 7   & 7   \\
Did you find the system recommended factor for improvement easy to use?                                       & 6   & 4   & 4   & 4   & 4   & 4   & 6   & 5   \\
Was comparing a technique in the analyzed speech to the speeches in our collection useful?                    & 7   & 6   & 4   & 7   & 4   & 6   & 7   & 7   \\
Was comparing a technique in the analyzed speech to the speeches in our collection easy to use?               & 6   & 4   & 5   & 4   & 4   & 5   & 7   & 6   \\
Was it useful to find speeches that had very similar or different use of techniques than the analyzed speech? & 7   & 5   & 4   & 7   & 5   & 6   & 7   & 7   \\
Was it easy to find speeches that had very similar or different use of techniques than the analyzed speech?   & 7   & 5   & 6   & 4   & 6   & 6   & 7   & 7   \\
Was it useful to find speeches that had very similar or different speech content than the analyzed speech?    & 6   & 6   & 5   & 7   & 4   & 6   & 7   & 6   \\
Was it easy to find speeches that had very similar or different speech content than the analyzed speech?      & 7   & 3   & 3   & 4   & 4   & 6   & 6   & 4   \\
Was it useful to observe the use of body language directly in the video feed?                                 & 7   & 5   & 6   & 7   & 3   & 6   & 7   & 5   \\
Was it easy to observe the use of body language directly in the video feed?                                   & 7   & 6   & 5   & 5   & 4   & 6   & 6   & 7   \\
Was it useful to find places in your speech to improve?                                                       & 7   & 5   & 5   & 7   & 4   & 6   & 7   & 7   \\
Was it easy to find places in your speech to improve?                                                         & 7   & 4   & 5   & 4   & 4   & 6   & 7   & 4   \\
Was it useful to use the script integrated with effectiveness data?                                           & 6   & 5   & 5   & 7   & 4   & 7   & 7   & 7   \\
Was it easy to use the script integrated with effectiveness data?                                             & 7   & 5   & 4   & 6   & 5   & 5   & 7   & 7   \\
Was it useful to understand how speech techniques change over time?                                           & 7   & 5   & 6   & 6   & 5   & 6   & 7   & 7   \\
Was it easy to understand how speech techniques change over time?                                             & 7   & 5   & 6   & 3   & 5   & 5   & 7   & 7   \\
Was it useful to summarize a speech with the methods given in our system?                                     & 7   & 6   & 5   & 7   & 6   & 7   & 7   & 7   \\
Was it easy to understand the summary of a speech with the methods given in our system?                       & 7   & 6   & 5   & 4   & 6   & 5   & 7   & 6   
\end{tblr}}
\end{table*}

\bibliographystyle{abbrv-doi-hyperref}

\bibliography{template}